\documentclass[10pt,journal,compsoc]{IEEEtran}
\usepackage{xcolor}
\usepackage{amsmath,graphbox}
\usepackage{amsfonts}
\usepackage{amssymb}
\usepackage{algorithm}[noend]
\usepackage{algpseudocode}
\usepackage{tikz}
\usepackage{multirow}
\usepackage{bbm}
\DeclareMathOperator*{\argmax}{arg\,max}
\DeclareMathOperator*{\argmin}{arg\,min}
\usetikzlibrary{shapes.geometric, arrows}
\usepackage{subfigure}
\usepackage{booktabs} 
\usepackage{thm-restate}
\usepackage{subfigure}
\usepackage{booktabs} 
\usepackage{thm-restate}

\usepackage{xcolor}
\usepackage[pagebackref=false,breaklinks=true,bookmarks=true]{hyperref}

\usepackage{cleveref}

\usepackage{dsfont}
\usepackage{amssymb}
\usepackage{makecell}

\def\G{{\mathcal{G}}}
\def\M{{\mathcal{M}}}

\usepackage[nocompress]{cite}

\ifCLASSINFOpdf
\else
\fi
\usepackage{array}
\begin{document}
%
\title{Generating 2D and 3D Master Faces for Dictionary Attacks with a Network-Assisted Latent Space Evolution}
%
%
%
%

\author{\parbox{16cm}{\centering
    {\large Tomer Friedlander$^{1*}$, Ron Shmelkin$^{2*}$, Lior Wolf$^{2}$}\\
    {\normalsize
    $^1$The School of Electrical Engineering, Tel Aviv University\\
    $^2$The Blavatnik School of Computer Science, Tel Aviv University\\
}}
    \thanks{*Equal contribution}
}

%
%

\markboth{Journal of IEEE TRANSACTIONS ON BIOMETRICS, BEHAVIOR, AND IDENTITY SCIENCE}%
{Journal of IEEE TRANSACTIONS ON BIOMETRICS, BEHAVIOR, AND IDENTITY SCIENCE}
\IEEEtitleabstractindextext{%
\begin{abstract}
A master face is a face image that passes face-based identity authentication for a high percentage of the population. These faces can be used to impersonate, with a high probability of success, any user, without having access to any user information. We optimize these faces for 2D and 3D face verification models, by using an evolutionary algorithm in the latent embedding space of the StyleGAN face generator. For 2D face verification, multiple evolutionary strategies are compared, and we propose a novel approach that employs a neural network to direct the search toward promising samples, without adding fitness evaluations. The results we present demonstrate that it is possible to obtain {a considerable coverage of the identities in the LFW or RFW datasets} with less than 10 master faces, for six leading deep face recognition systems. 
In 3D, we generate faces using the 2D StyleGAN2 generator and predict a 3D structure using a deep 3D face reconstruction network. When employing two different 3D face recognition systems, we are able to obtain a coverage of 40\%-50\%. {Additionally, we present the generation of paired 2D RGB and 3D master faces, which simultaneously match 2D and 3D models with high impersonation rates.}
\end{abstract}

}

\maketitle

\IEEEdisplaynontitleabstractindextext

%

\IEEEraisesectionheading{\section{Introduction}\label{sec:introduction}}

\IEEEPARstart{I}{n} dictionary attacks, one attempts to pass an authentication system by
trying multiple inputs. In real-world biometric systems, one can typically try only a handful of inputs before being blocked. However, the matching in biometrics is not exact, and the space of biometric data is not uniformly distributed. This may suggest that with a handful of samples one can cover a high percentage of the population.
{Our results show that for both 2D and 3D face recognition, there are face images that would be authenticated successfully for a high percentage of users in a given dataset. This is demonstrated using state-of-the-art face recognition systems and acceptable matching thresholds. In some cases, a single 2D face can cover more than 20\% of the identities in LFW~\cite{ds:LFWTech} and a 3D master face covers more than 32\% of the identities in the Texas 3D dataset~\cite{gupta2010texas}. Following previous work on fingerprints~\cite{roy2017masterprint}, we term such faces ``master faces'', due to the analogy to master keys.

The process of master face generation employs a realistic face generator network, in our case StyleGAN~\cite{styleGan, Karras2019stylegan2} for generating 2D faces. In 3D, a deep 3D reconstruction network is applied on the output of the 2D generator to obtain the corresponding 3D structure. Since the objective function, i.e., the number of identities similar enough to the face image we optimize, is non-differentiable, we use black box optimization methods. As expected, we find that methods tailored to high-dimensional data outperform other methods. We then propose a novel method, in which a trained neural network is used to predict the competitiveness of a given sample.
}

When attempting to cover an even larger number of faces, we advocate a greedy method in which one repeats the master face generation process sequentially, each time attempting to cover the identities that were not covered by previously generated faces. Using such a greedy process, we obtain a coverage of 6\%-60\% with nine images (the focus on nine images arises from a different experiment, in which the samples in the latent space are clustered).

The experiments are conducted using six different deep face recognition models for the 2D scenario and two different deep face recognition models for the 3D systems, each with its own processing pipeline, training dataset, objective, and similarity measure. For the 2D evaluations on the LFW dataset~\cite{ds:LFWTech}, a conservative similarity threshold based on obtaining a standard FAR value of 0.001 is used, or, when available, the threshold prescribed by the method itself. For the 2D evaluations on the raccially-diverse RFW dataset~\cite{Wang_2019_ICCV} and also in the 3D experiments, the unique threshold that balances FAR and FRR is used.

Overall, our results indicate that performing a dictionary attack on face authentication systems is feasible, with high rates of success. This is demonstrated for both 2D and 3D, employing multiple face representations, and explored with several state-of-the-art optimization methods.

\section{Related work}
Face recognition is a dominant biometric modality, with applications in user authentication, surveillance, automatic photo tagging, etc. The field greatly benefited from the availability of web-scale datasets. The same datasets, coupled with the relatively stable structure of the face have also contributed to the development of effective methods for synthesizing random face images.

\label{sec:preliminaries}
\subsection{Face Verification}
\label{subsec:Face Recognition Systems}

{Face Verification is the task of comparing two face images and determining whether they belong to the same subject. In the past few years, deep neural network (DNN) approaches have dominated this field. These DNNs learn to embed faces in some representation space, where similarity metrics can be used to perform the verification task. FaceNet~\cite{fd:Facenet} is trained using the triplet loss and utilizes the euclidean metric. SphereFace~\cite{fd:sphereface} employs a novel Angular-Softmax loss as an improved metric.} CosFace~\cite{wang2018cosface} proposes an additive cosine loss, while ArcFace~\cite{deng2019arcface} employs an additive angular loss. MagFace~\cite{meng2021magface} proposes a magnitude-aware margin for the ArcFace loss~\cite{deng2019arcface} that allows the model to learn an embedding, whose magnitude is able to measure the given face's quality. MagFace seeks to pull high quality face samples close to the class centers, while it pushes low quality samples away. Recently, ElasticFace~\cite{boutros2022elasticface} introduced a novel elastic loss term that relaxes the fixed penalty margin that is frequently used. Two variants, ElasticFace-Arc and ElasticFace-Cos, are created by integrating this novel randomized margin penalty with the loss terms of either ArcFace~\cite{deng2019arcface} or CosFace~\cite{wang2018cosface}.

Face verification is often assessed by considering the False Acceptance Rate (FAR) and False Rejection Rate (FRR). FAR is the percentage of embedding representations in which different subjects were incorrectly authorized as the same person, FRR is the percentage of embedding representations in which embedding representations belonging to the same subject were incorrectly rejected. There is a trade-off between these two metrics, which is balanced by a matching distance threshold $\theta$. Such a threshold is usually defined by defining a targeted FAR or by defining a unique threshold of interest, for which the Equal Error Rate (EER) is obtained, i.e. the FAR is equal to the FRR, thus balancing the two rates.

2D face recognition systems were the first to be researched and adopted by the industry, mainly due to the availability of 2D face images. However, recent advances in sensor technologies have enabled highly accurate 3D face recognition systems~\cite{zhou20183d,jing20213d}. The general steps of 3D face recognition methods are similar to the 2D case, consisting of face detection, embedding extraction, and embedding matching. However, the features used for 3D faces differ in their semantic representation from features used for 2D. While 2D models might consider features such as age, race, skin color, etc., 3D models mainly consider geometric information of facial features. Thus, under certain conditions, 3D models can even outperform the SoTA 2D face recognition models \cite{bowyer2006survey,FaceRec3D}. The extraction of these features is more complicated, usually requiring special scanners and formats. The three most common formats for 3D face representation are: 1) Depth Images, 2) Range Images, 3) Mesh. The use of geometric features makes the method sensitive to pose, expression and illumination of the face. As a result, in recent years, 3D face recognition research has been directed at the creation of methods that are insensitive to these factors \cite{zhou20183d}.

Our experiments are conducted on two 3D face recognition networks: Kim et al. \cite{kim2017deep} and FR3DNet \cite{FaceRec3D} . Kim et al. \cite{kim2017deep} proposed a deep convolutional neural network by fine-tuning VGG-Face \cite{parkhi2015deep}, which had been pre-trained on 2D RGB images. For its training, 2D depth images projected from augmented 3D face scans were used. The method is reported to perform well on several extensively used 3D face recognition datasets \cite{Bosphorus,BU3DFE,3DTEC}.

FR3DNet \cite{FaceRec3D} is a deep convolutional model that was specifically designed for the task of 3D face recognition by receiving a three-channel input: the depth image fitted to the 3D scan, and the azimuth and elevation angles of the normal to the 3D surface. FR3DNet was trained from scratch on 3.1 million facial scans of 100k identities. In addition, it is reported to achieve leading results on a test dataset of 31,860 3D scans of 1,853 identities. This test dataset, which merges the most challenging existing public 3D datasets \cite{FRGCv2,BU3DFE,Bosphorus,GavabDB,gupta2010texas,BU4DFE,CASIA3D,UMBDB,3DTEC,ND2006}, is the largest 3D face dataset and has been reported for evaluating 3D face recognition models.

\subsection{Face Image Generation using GAN}
The Generative Adversarial Network (GAN)~\cite{GAN} is a machine learning framework in which two neural networks, the generator $G$ and the discriminator $D$, are trained in the form of a min-max game. Given the dataset distribution $X$ and a latent input vector $z\sim Z$, the goal of $G$ is to generate samples $G(z)\sim X$. Given a data sample $x$, $D$ is trained to determine whether $x\sim X$ or $x\sim G(z)$, where $z\sim Z$. On the other hand, $G$ is trained to create samples for which $D$ fails to discriminate between the two. The quality and resolution of data samples generated using GANs have been constantly improving. One of the most notable lines of work is that of StyleGAN Face Image Generation~\cite{styleGan}.

StyleGan~\cite{styleGan} applies a mapping neural network $f: Z\rightarrow W$ to convert the latent vector $z$ into a more disentangled latent representation $w$ that separates content and style. It is then fed to each convolutional layer of the generator through adaptive instance normalization (AdaIN)~\cite{AdaIn}. This allows better control of the image synthesis process and results in the generation of high-quality and more detailed images. {In order to reduce the presence of artifacts in the generated face images, improvements were introduced to the training procedure and the network architecture in a subsequent work, StyleGan-V2~\cite{Karras2019stylegan2}}.

\subsection{Master sample attacks}
\label{sec:previous_work}
{
Roy et al.~\cite{roy2017masterprint} generate fingerprint templates that can be matched to a large number of users' fingerprints, without any knowledge of the actual user. This generation does not produce an actual image, which would be required for the actual match. To produce such an image, Bontrager et al.~\cite{deepmasterprints} used a deep neural network.

Our work performs a similar task for faces. In comparison to fingerprints, faces are characterized by a larger latent space, and we develop an optimizer that is more suitable for large dimensions. Through an extensive set of experiments, we explore various state-of-the-art solutions and demonstrate the effectiveness of our method.

Similar to our work, Nguyen et al.~\cite{nguyen2020generating} generate master faces using StyleGAN, but with a different evolutionary strategy, which is used as one of our baselines (CMA-ES). Subsequently, an extensive analysis of master face generation, examines its transferability and the combination of different models and datasets, and attempts to set out conditions for the strong master face~\cite{nguyen2021master}.

Our concurrent work~\cite{ourMasterFace} presents Master Face generation for 2D face verification systems, evaluates various evolutionary algorithms, and presents a Network-Assisted Latent Space Evolution. It  explores adding a  success predictor model to the LM-MA-ES evolutionary algorithm, and presents a greedy algorithm for generating a minimal number of different master faces in order to be faultily authenticated with a maximal number of subjects in the targeted dataset.

}

Another work~\cite{gernot2022biometric} formalizes the problem of master key generation for cancellable biometric databases. This is done for a face dataset, a fingerprint dataset and an ECG dataset. 

This current manuscript extends~\cite{ourMasterFace}. {For 2D (image) faces, it generates master faces for more recent face recognition models (\cite{meng2021magface} and two different versions of \cite{boutros2022elasticface}) and adds evaluations on the RFW dataset~\cite{Wang_2019_ICCV,wang2021meta,wang2019skewness,wang2018deep} which is more racially-diverse than LFW.} In addition, this extension includes the treatment of 3D face recognition and the generation of master faces that simultaneously match 2D and 3D models with high impersonation rates.

\subsection{Evolution Strategies}\label{subsec:evolution_strategies}
Covariance Matrix Adaptation Evolution Strategy (CMA-ES)~\cite{hansen2003reducing} is a type of iterative evolutionary algorithm widely used for solving non-convex continuous optimization problems without using gradient information. It is considered one of the most powerful stochastic numerical optimizers for solving difficult black-box problems~\cite{varelas2018comparative}.

At each iteration, CMA-ES generates a population of $\lambda$ candidate solutions, by sampling a multivariate Gaussian distribution, whose mean vector and covariance matrix were estimated in the previous iterations. The $\mu$ best fitting candidates in terms of the objective function are selected from the current population and are used to adjust the last estimate of the model's learnable parameters.

The quadratic time and space complexity of CMA-ES limits its applicability for high-dimensional problems~\cite{varelas2018comparative} and in higher dimensions its performance degrades significantly~\cite{omidvar2010comparative}. 
Limited-Memory Matrix Adaptation Evolution Strategy (LM-MA-ES)~\cite{loshchilov2017limited} 
reduces the time and storage complexity to $\mathcal{O}(n\log{n})$. This algorithm is reported to perform well on high-dimensional variants of well-established benchmark problems. This makes LM-MA-ES an appropriate choice for solving the high-dimensional black-box optimization problem in the current work.

Evolutionary algorithms assisted by an additional machine learning model have been presented in the literature, e.g., surrogate-assisted CMA-ES algorithms \cite{pitra2017overview}. { Some models train a regression model online with the evolutionary algorithm in order to learn the fitness function of the black-box problem \cite{surr_cmaes}. Instead of the expensive computation of the fitness function itself for all candidates, the fitness scores of some candidates are predicted by this regression model, thereby reducing the number of evaluation calls. Models can be assisted by the surrogate for another purpose: predicting whether a given candidate is going to be competitive in terms of its fitness score and evaluating only such candidates. ACM-ES \cite{loshchilov2010comparison}, which uses a comparison-based surrogate instead of a regression model, is an example of such models.
In our work, we train a neural classifier coupled with the evolutionary algorithm in order to predict a given candidate's probability to be fitter (lower fitness) relative to candidates generated in the last few iterations, without evaluating its fitness score explicitly.
}

\section{METHODOLOGY}
\label{sec:method}
Given a dataset $D=\{x\in \mathbbm{R}^{w \times h\times c}\}$ containing face images (each of size $w\times h$ and with $c$ channels) with a single image per subject, a deep convolutional face embedding model $\M(x)\in \mathbbm{R}^d$ and a matching threshold $\theta$, we define the \textbf{Master Face $\mathbf{x_{mf}}$} as follows:
\begin{equation}\label{eq:x_mf}
\mathbf{x_{mf}} = \argmax_{x_{mf}} \sum_{x \in D}f(\M(x_{mf}),\M(x)),\theta)  
\end{equation}
The binary function $f:\mathbbm{R}^{2d+1} \rightarrow \{0,1\}$ compares two embedding vectors and assigns a value of $1$ if the embeddings are considered similar enough according to a given similarity metric and a predefined recognition threshold ($\theta)$.
We note that our method searches for an optimal face image $x_{mf}$ and for an optimal embedding 
\begin{equation}\label{eq:c_mf}
  c = \argmax_{c} \sum_{x \in D}f(c,\M(x),\theta)  
\end{equation}
As we show in Sec.~\ref{subsec:cent_inv}, it is not possible to invert $\M$ effectively and obtain a 
$\M^{-1}(c)\in \mathbbm{R}^{w \times h\times c}$ that achieves both high visual quality and high coverage. 

Instead, we suggest optimizing the Face Generator's latent vector $z$ based on its matching score, in order to find a better representation in the image space.
\begin{equation}\label{eq:z_optpre}
  \mathbf{z_{opt}} = \argmax_{z} \sum_{x \in D}f(\M(\G(z)),\M(x)),\theta)
\end{equation}
{
By summing over the targeted dataset, we count the number of objectives that were successfully authenticated by the suggested master face, such that $0 \le \sum_{x \in D}f(\M(\G(z)),\M(x),\theta) \le n$.
By convention, evolutionary algorithms minimize the score instead of maximizing it. Therefore, the above matching score is modified by subtracting it from the total number of face images in the training set $n=|D|$. Moreover, the score is normalized to be in the $[0,1]$ range. 

\begin{equation}\label{eq:z_opt}
  \mathbf{z_{opt}} = \argmin_{z}(1- \frac{1}{n}\sum_{x \in D}f(\M(\G(z)),\M(x),\theta))
\end{equation}
}
\begin{figure}[t]
  \centering
    \includegraphics[scale=0.35]{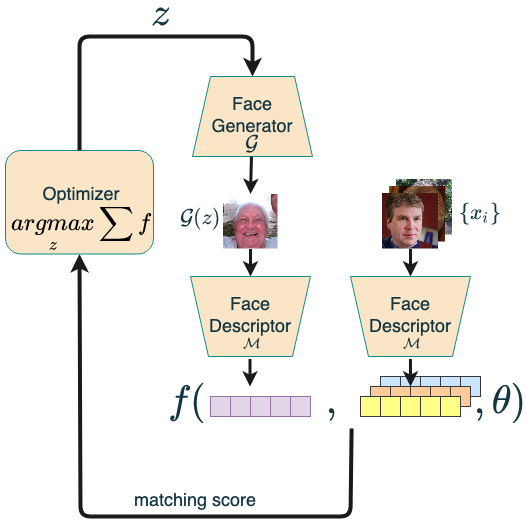}
\caption{An overview of the generation process for finding a master face.}
\label{fig:alg_flow}
\end{figure}
An overview of the method is given in Fig.~\ref{fig:alg_flow}. Initially, the dataset $D$ is transformed to the embedding space. Denote the embedded dataset by $\widehat{D}=\{\M(x):~\forall x \in D\}$. An evolutionary algorithm is used to find the optimal latent vector $\mathbf{z_{opt}}$ that solves the optimization problem defined in Eq.~\ref{eq:z_opt}. At each iteration of the optimization algorithm, a set of candidate solutions is generated and evaluated by the fitness function. Given a candidate latent vector $z$, an image corresponding to $z$ is generated by applying the face generator $\G$ on $z$. The face is then extracted from the generated image $\G(z)$ and is embedded using the face description model $\M$.

\subsection{The evolutionary algorithm}\label{subsec:opt_imp}
\begin{figure*}[t]
\centering
\begin{minipage}[b]{0.425\linewidth}
    \centering
    \includegraphics[scale=0.29]{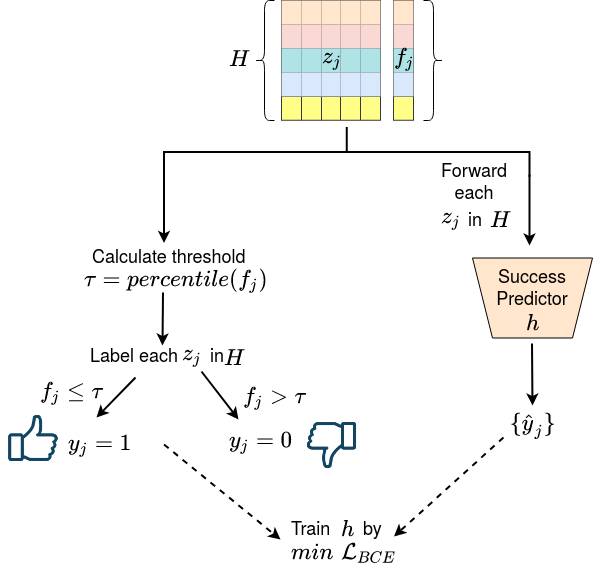}\label{fig:training_predictor}
    \centerline{(a)}
\end{minipage}
\begin{minipage}[b]{0.425\linewidth}
    \centering
    \includegraphics[scale=0.29]{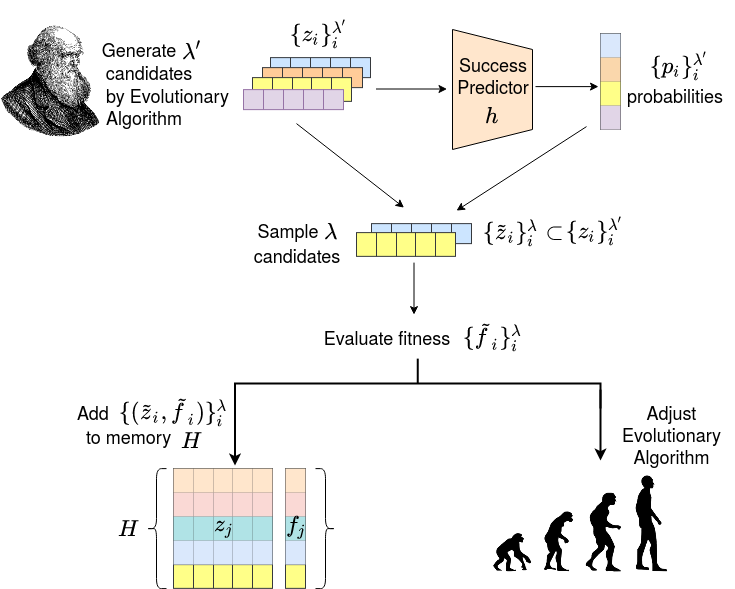}     \label{fig:filtering_by_predictpr}
     \centerline{(b)}
\end{minipage}
\caption{(a) Training the success predictor using samples stored in memory $H$. (b) Filtering generated candidates by the success predictor and updating $H$.}
\label{fig:success_predictor_flow}
\end{figure*}

We present an evolutionary algorithm that is coupled with a neural predictor. The latter estimates the probability of a given candidate being {fitter (lower fitness score)} than the p-percentile of candidates generated during the last few iterations. {These estimated probabilities are used to obtain a set of promising candidates for the purpose of enhancing latent space optimization.} 

The LM-MA-ES evolutionary algorithm, which is known to perform well on high-dimensional black box optimization problems, is selected as the baseline method for the optimization problem in Eq.~\ref{eq:z_opt}. At each iteration, the original variant of LM-MA-ES generates a population of $\lambda$ candidate solutions and the $\mu$ fittest candidates are selected for adjusting the parameters of the model. Since the candidates are generated randomly, by sampling the probability density learned by the LM-MA-ES model, some candidates may be unsuccessful in terms of their fitness score. In order to obtain a set of successful candidates with a higher probability, we suggest training a neural classifier for predicting {the probability of a given candidate being fitter than the p-percentile of stored samples from the few last iterations.} We use this classifier, named the Success Predictor, to filter a larger generated set to the candidates more likely to be promising, without having to evaluate their fitness scores explicitly.

Let $h$ be a neural classifier, which receives a vector $z$ as input. The training process of classifier $h$ is described in Fig. \ref{fig:success_predictor_flow}(a) and is performed online, at the end of each iteration of the evolutionary algorithm. The training samples are candidates generated during the last few iterations of the evolutionary algorithm, which are stored in a finite memory $H$ with a capacity of $C$ samples. Each new evaluated candidate and its fitness score are added to the memory $H$. {When the memory $H$ becomes overloaded, surplus samples are removed from it randomly, preserving the best candidate.} After each update of the memory $H$, each sample $z_i$ in $H$ is defined to be a successful candidate (labeled +1) if its fitness score is better than the p-percentile of the fitness scores of all samples stored in the memory. Otherwise, it is considered an unsuccessful candidate (labeled 0). Let $(z_i, y_i) \in H$ be the i-th sample in the memory and its corresponding label. The classifier is trained to classify the class of the training samples successfully, by minimizing the binary cross entropy loss, $\mathcal{L}_{BCE}$ between the predicted class and the the true class.

The classifier $h$ is used in the inference mode during the generation step of the evolutionary algorithm, as described in Fig. \ref{fig:success_predictor_flow}(b). Instead of generating only $\lambda$ candidates, $\lambda' > \lambda$ candidates are generated. Prior to evaluating the candidates' fitness values, each candidate is forwarded through the classifier $h$ in order to obtain a {score representing the probability of it belonging to the successful class of samples. The scores of all $\lambda'$ candidates are concatenated to a single vector, which is afterwards converted to a probability vector by applying the Softmax operator.} Following this, only $\lambda$ candidates are selected for evaluation, by sampling them out of the total $ \lambda'$ candidates according to the obtained probability scores. Other steps of the evolutionary algorithm remain unchanged and are performed on these $\lambda$ candidates, in particular, the fittest $\mu$ candidates chosen from this pre-filtered set.

{After evaluating the fitness scores of the new filtered candidates, we can retrospectively determine whether these candidates {were correctly classified by the classifier $h$ earlier. In particular, we determine if a candidate that obtained a prediction score higher (lower) than 0.5 by $h$, is indeed (less fit) fitter than the p-percentile of the samples in the memory $H$.} The performance of $h$ is tested by calculating the average prediction accuracy for these new candidates. If this average accuracy drops below a predefined threshold $\tau_{acc}$ for a predefined number of $T$ iterations, the learnable parameters of $h$ are re-initialized at random. Moreover, we use a predefined warm-up period of iterations at the first generations of the evolutionary algorithm for which the classifier is only trained, but is not used for filtering new candidates.}

\subsection{Dataset coverage}\label{subsec:coverage}
Given a set of face images $D$, we strive to find a minimal set of master face images $S=\{x\in \mathbbm{R}^{w\times h\times x}\}$ such that for as many subjects $x\in D$ as possible, there is a least one $x'\in S$ such that $f(\M(x),\M(x'),\theta)$ is one.

A natural choice is to divide the embedding space into clusters, e.g., by using KMeans~\cite{kmeans}, and optimize each member of $S$ to cover a different cluster, e.g., by considering the center of each cluster. However, as shown in~\ref{subsec:cent_inv} {, for the 2D face images, } it is difficult to invert the cluster centroid point to the image space. We therefore propose a greedy approach Alg.~\ref{alg:greedy} to find such a set of master face images. After each iteration, images that were incorrectly authorized {with the generated master face of the current iteration} are removed from the dataset $D$, and the next search iteration is performed on the updated dataset. {\color{black}Therefore, the current generated master face might cover some face images that were already covered by previously generated master faces, but the opposite cannot occur. However, such an intersection does not result in counting covered face images twice in the total coverage percentage calculation, since the current coverage iteration is performed on the reduced dataset.} 

The method uses the function $find\_matched$, which given a face image $x_{mf}^i$ and the dataset $D$, returns the set of faces from $D$ that are incorrectly authorized with $x_{mf}^i$. We set the limit of the number of iterations in Alg.~\ref{alg:greedy} to be 
the number of clusters that cover most of the $\widehat{D}$, as presented in Sec.~\ref{subsec:cent_inv}.
\begin{algorithm}[t]
  \begin{algorithmic}[1]
    \Function{find coverage}{$\G, D, \M, \theta, max\_iter $}
    \State $center\_imgs \gets []$
    \For {$i = 1..{ max\_iter}$}
        \State $x_{mf}^i \gets\textrm{master face generation}(\G, D, \M, \theta)$ 
        \State $center\_imgs \gets center\_imgs \bigcup x_{mf}^i$ 
        \State $D \gets D \backslash\ find\_matched(D,x_{mf}^i)$
    \EndFor
    \State\Return $center\_imgs$
\EndFunction
    \end{algorithmic}
\caption{$greedy$-Coverage Search}\label{alg:greedy}
\end{algorithm}

\section{Experiments - 2D Scenario}
\label{sec:experiments_2d}

We evaluated our method with six different CNN-based face descriptors: Dlib~\cite{fd:dlib09}, FaceNet~\cite{fd:Facenet}, SphereFace~\cite{fd:sphereface},  MagFace~\cite{meng2021magface} and two versions of ElasticFace~\cite{boutros2022elasticface} that are trained for minimizing different losses (ArcFace~\cite{deng2019arcface} and CosFace~\cite{wang2018cosface}). Each face descriptor is equipped with its own combination of an architecture, a similarity metric and a loss function, thus providing additional validation of our method.

The Dlib face descriptor ~\cite{fd:dlib09} employs a ResNet with 29 convolutional layers. This architecture is ResNet-34~\cite{he2015deep} with a reduced number of layers, in which the number of filters per layer is halved. The reported LFW accuracy is $0.9938$. The model is trained on the face scrub~\cite{ds:scrub}, VGG~\cite{ds:vgg} and additional face images scraped from the internet. The FaceNet implementation employs an Inception-ResNetV1~\cite{resnetv1}. It achieves $0.9905$ accuracy on the LFW face verification benchmark. The SphereFace face descriptor is implemented as a deep neural network with 20 convolutional layers. An accuracy of $0.9922$ is measured on the LFW face verification benchmark. Both FaceNet and SphereFace were trained on the CasiaWebface~\cite{ds:casia} dataset. The MagFace~\cite{meng2021magface}, ElasticFace-Arc~\cite{boutros2022elasticface} and Elastic Face-Cos~\cite{boutros2022elasticface} implementations employ an iResNet100~\cite{duta2021improved} architecture and  were trained on the MS1MV2 dataset~\cite{guo2016ms,deng2019arcface}. The reported LFW accuracy of MagFace, ElasticFace-Arc and ElasticFace-Cos is $0.9983$, $0.9980$ and $0.9982$, respectively.

The face detection and alignment procecsses of Dlib uses Dlib's dedicated detector ~\cite{dlib_fd}. Dlib embeds the face images in $\mathbbm{R}^{128}$ and employs the Euclidean distance. For all other model, the face regions are extracted and aligned using MTCNN~\cite{mtcnn}, the similarity of embeddings is measured by the cosine distance, and the embedding dimension is 512. 

The experiments that were done in our previous work~\cite{ourMasterFace} use the StyleGAN model~\cite{styleGan} pre-trained with the FFHQ~\cite{styleGan} dataset as the face generator, $\G$. In the new experiments of this current extension (recent face descriptors MagFace, ElasticFace-Arc and ElasticFace-Cos, and all the experiments on the RFW dataset) we use the newer StyleGAN2~\cite{Karras2019stylegan2} model trained on FFHQ~\cite{styleGan} in order to generate even better quality master faces. In both cases cases, $\G: \mathbbm{R}^{512} \rightarrow \mathbbm{R}^{1024\times1024\times3}$.

The architecture of Success Predictor $h$ is a feed-forward neural network with three fully connected layers, whose output dimensions are 256, 128 and 1 neurons, respectively. The first two hidden layers and the output layer are followed by the ELU \cite{clevert2015fast} activation function and a Sigmoid, respectively. The first hidden layer uses the BatchNorm regularization layer \cite{ioffe2015batch}, prior to the activation function. The network is trained using the ADAM \cite{kingma2014adam} optimizer with a learning rate of $0.001$ on mini-batches of size 32. The population size $\lambda$ is set to $22$ according to the calculation in \cite{loshchilov2017limited} and $\lambda'$ is set to 1,000 candidates. The threshold defining the promising class is set to the 5th percentile. The capacity $C$ of the memory $H$ is set to 5000 samples. The parameters $\tau_{acc}$ and $T$ are set to 0.6 and 20, respectively. The warm-up period of $h$ is set to the first 5\% of the evolutionary algorithm iterations.

The method is evaluated on the LFW~\cite{ds:LFWTech} dataset and on {the RFW dataset~\cite{Wang_2019_ICCV,wang2021meta,wang2019skewness,wang2018deep}}, the latter is meant to provide a racially-balanced dataset. One image per subject is used.
Note that, since we compare a new generated face to the set of different subjects, we measure the matching score as Mean Set Coverage (MSC). 
$$MSC=\frac{\# \textrm{of incorrect authorization }}{|D|} * 100$$

{For the experiments on LFW, we use Dlib's predefined recognition threshold ($\theta$) of 0.6, while for all other models we choose the threshold that preserves $FAR\sim0.001$. Regarding the experiments on RFW, we noticed that the FRR of some of the face recognition models is too large (e.g. $\sim0.3$ by SphereFace) on the training subset at the FAR of $0.001$. Therefore, we decided to employ the Equal Error Rate (EER) trade-off point, i.e. the recognition threshold ($\theta$) is selected approximately at the equilibrium of the FAR and the FRR measurements. The measured EER on the training set in the RFW experiments is $\sim0.04$ for Dlib, FaceNet and SphereFace, $\sim0.004$ for MagFace and $\sim0.003$ for ElasticFace-Arc and ElasticFace-Cos.}

\subsection{Cluster centroid inversion}\label{subsec:cent_inv}
\begin{figure}[t]
\centering
\begin{tabular}{@{~}c@{~}c@{~}c@{~}c@{~}c@{~}}
    \includegraphics[width=1.3cm]{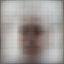} & \includegraphics[width=1.3cm]{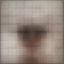} & \includegraphics[width=1.3cm]{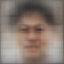} &      \includegraphics[width=1.3cm]{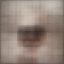} &   
    \includegraphics[width=1.3cm]{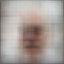} \\ \includegraphics[width=1.3cm]{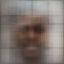} & \includegraphics[width=1.3cm]{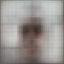} &      \includegraphics[width=1.3cm]{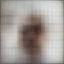} &   
    \includegraphics[width=1.3cm]{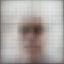} \\
\end{tabular}
\caption{Centroids in latent space, converted to image space}\label{fig:cent_images}
\end{figure}
We first explore the alternative method of clustering the data in the embedding space and then trying to convert the prototypes found into images. This also provides an estimate of the number of target master faces one can use in an ideal case where a dictionary attack can be made with embeddings instead of real-world faces.

In this experiment, we focus on the Dlib face descriptor. An embedding dataset $\widehat{D}$ is created and then clustered using KMeans~\cite{kmeans}. With nine clusters $\sim 91.63\%$ of $\widehat{D}$ was covered, i.e. we were able to find nine center points (cluster centroids) in the embedding space such that over 90\% of the samples in $\widehat{D}$ were within the Euclidean distance of less than $\theta=0.6$ from at least one of these centroids.

Since the actual master-based dictionary attack requires using images, we trained a neural network to generate a face image from Dlib's embedding representation. Specifically, we trained a neural network face generator $G:\mathbbm{R}^{128}\rightarrow\mathbbm{R}^{64x64x3}$ with five layers. The layers consist of one linear layer and four de-convolutional layers, similarly to the generator architecture presented in DC-GAN~\cite{dcgan}. The generator $G$ was trained on Dlib's embedding representation of the FFHQ\cite{styleGan} dataset. It is trained to minimize the MSE loss between the generated image and the original one.
\begin{equation}
\mathcal{L}_{MSE} = \frac{1}{|D_{FFHQ}|}\sum_{x\in D_{FFHQ}}(x - G(\M(x)))^2 
\end{equation}
While $G$ performs well on embeddings of the real image, the visual quality in the case of cluster centroids is unsatisfactory, see Fig.~\ref{fig:cent_images}. Moreover, only two out of the nine generated faces are detected as faces by the Dlib detector. These generated faces are also ineffective as master faces and the highest MSC score of any of the nine is only $\sim2\%$.

\subsection{Experiments for one Master Face image}\label{subsec:gf_eval}
The generic scheme we present for optimizing a single image, as depicted in Fig.~\ref{fig:alg_flow}, is employed for recovering a single Master Face image on the LFW dataset. We evaluated our approach on the predefined split $|D_{train}|/|D_{test}|=4038/1711$. For each face descriptor, each black-box optimization method was trained on $D_{train}$ for five runs. Each run differs in its initial random seed. Out of the face images obtained from all five runs, the one that achieved the highest MSC score on the training dataset was chosen to be reported as the master face obtained for this face descriptor on the training dataset. This way, we eliminate some of the sensitivity of such optimization methods to the random seed, without using any test data.

To enable a fair comparison, all algorithms were trained for an equal number of fitness function calls (26400), with the same set of five seeds. The number of fitness function calls was chosen based on the observation that longer training processes resulted in negligible improvement. After this selection process, each master face was then evaluated on the test dataset, $D_{test}$. 

We compare the performance of our method, which is denoted by LM-MA-ES + Success Predictor with the following baselines: (i) A random Search algorithm that was used to set the baseline results. We used the version implemented in the Nevergrad package \cite{nevergrad}. (ii) LM-MA-ES~\cite{loshchilov2017limited}, the high-dimensional variant of CMA-ES~\cite{hansen2003reducing}, which our method is based on. (iii) For completeness, the original CMA-ES algorithm, despite it being unsuitable for high dimensions. We use the implementation from the pycma package~\cite{hansen2019pycma}.

Differential Evolution (DE) is another highly successful family of evolutionary algorithms. In addition to the original (iv) DE \cite{storn1997differential}, we also compare our method with the newer variants (v) LSHADE-RSP~\cite{stanovov2018lshade} (second place in IEEE CEC'2018) and (vi) IMODE~\cite{sallam2020improved} (first place in IEEE CEC'2020).

(vii) NGOpt~\cite{liu2020versatile} is an algorithm that automatically selects the right evolutionary algorithm to be trained out of a set of several algorithms, according to the properties of the optimization problem. NGOpt is implemented in the Nevergrad package.
{ (viii) ACM-ES \cite{loshchilov2010comparison} is a surrogate-assisted CMA-ES variant with a comparison-based surrogate (ranking SVM). Similarly to our approach, it selects for evaluation only a subset of candidates predicted to be promising according to the surrogate model. The initial larger population and the warm-up period are set identically to the ones used in the model assisted by the Success Predictor. We use the implementation from the BOLeRo package \cite{bolero}. }
(ix) lq-CMA-ES \cite{surr_cmaes} is a surrogate-assisted CMA-ES variant assisted by a linear-quadratic regression model (LQ) { It uses the surrogate for decreasing the number of evaluations and not for predicting promising candidates.} We use the implementation from pycma \cite{hansen2019pycma}. (x) For the purpose of comparing our model more directly to another surrogate-assisted model, we add this LQ regression model to the LM-MA-ES algorithm and train it online as done in lq-CMA-ES. This variant is named LM-MA-ES + LQ-Filter. Similarly to our method, LM-MA-ES+LQ-Filter generates a larger set of $\lambda'>\lambda$ candidates. Their fitness score is then predicted by the LQ model and their success probability is estimated by applying the Softmax function on the negated predicted fitness scores. The estimated probabilities are used for sampling a smaller set of $\lambda$ candidates. To adjust the LM-MA-ES's parameters fairly, the actual fitness values of all selected candidates are evaluated, instead of using the values predicted by the LQ model.

\begin{table}[t]
\caption{MSC score (\%) of generated master faces for different optimization methods.}
\centering
\begin{tabular}{@{}l@{~~}c@{~}c@{~}c@{~}c@{~}c@{~}c@{}}
\toprule& 
\multicolumn{2}{c}{Dlib} &
\multicolumn{2}{c}{FaceNet} & 
\multicolumn{2}{c}{SphereFace}
\\
\cmidrule(lr){2-3}
\cmidrule(lr){4-5}
\cmidrule(lr){6-7}

Optimization Algorithm & Train & Test & Train & Test 
& Train & Test\\
\midrule
RandomSearch 
& $5.78$ & $4.73$
& $9.69$ & $9.94$
& $7.60$ & $6.78$ \\
DE 
& $7.08$ & $5.67$
& $12.29$ & $11.87$  
& $14.12$ & $13.68$ \\
LSHADE-RSP
& $10.09$ & $8.30$
& $15.90$ & $15.20$ 
& $15.73$ & $11.52$ \\
IMODE
& $7.70$ & $6.06$
& $12.36$ & $12.04$
& $11.64$ & $11.52$ \\
CMA-ES 
& $18.17$ & $16.32$
& $16.75$ & $16.43$ 
& $16.84$ & $15.49$ \\
NGOpt 
& $14.04$ & $12.71$ 
& $16.62$ & $15.20$ 
& $17.24$ & $16.37$\\
lq-CMA-ES
& $7.38$ & $6.12$
& $13.10$ & $12.57$ 
& $9.31$ & $9.82$ \\
ACM-ES
& $18.09$ & $17.61$
& $16.00$ & $14.50$ 
& $16.67$ & $16.84$\\
LM-MA-ES
& $18.27$ & $17.97$ 
& $17.12$ & $\mathbf{16.96}$ 
& $17.39$ & $16.96$\\
LM-MA-ES + LQ-Filter
& $18.87$ & $16.67$ 
& $16.60$ & $14.91$ 
& $\mathbf{17.79}$ & $17.42$\\
LM-MA-ES + Success Pred.
& $\mathbf{22.43}$ & $\mathbf{21.56}$ 
& $\mathbf{17.36}$ & $16.60$ 
& $17.51$ & $\mathbf{17.60}$\\ 
\bottomrule
\end{tabular}
\label{table:ea_comp}
\vspace{-5mm}
\end{table}

Table~\ref{table:ea_comp} presents a comparison between the different optimization algorithms for the master face generation task, in terms of the MSC score. It can be observed that our LM-MA-ES+Success Predictor achieved the highest result among all compared algorithms on the train set for two out of three face descriptors, when considering either the train set or the test set. When not leading, our LM-MA-ES assisted by the Success Predictor achieved the second best result. In comparison to the original LM-MA-ES, LM-MA-ES assisted by the Success Predictor achieved better results on all three training sets and on two out of three test sets. Moreover, the Success Predictor seems to improve the LM-MA-ES baseline algorithm  by a greater extent than the LQ regression model in all experiments, except for the training set of SphereFace. { In addition, LM-MA-ES assisted by the Success Predictor outperforms the comparison-based surrogate-assisted ACM-ES in all experiments.} An additional observation is that high MSC results on the training set are often preserved on the test set.
\begin{figure}[t]
\centering
\begin{tabular}{c@{~~~}c@{~~~}c@{~~}c@{~~}c@{~~}}
    & Dlib & FaceNet & SphereFace \\
    (a)&
     \includegraphics[align=c,align=c,width=1.4cm]{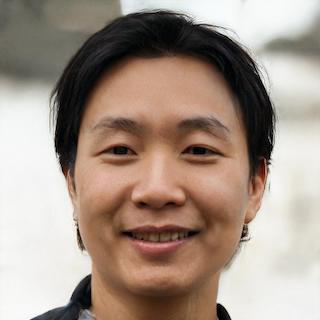} & \includegraphics[align=c,align=c,width=1.4cm]{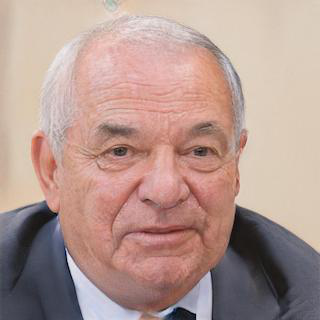} &  
    \includegraphics[align=c,align=c,width=1.4cm]{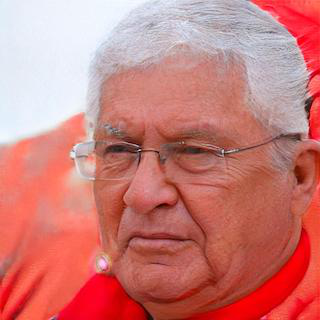} \\
     (b)&
     \includegraphics[align=c,width=1.4cm]{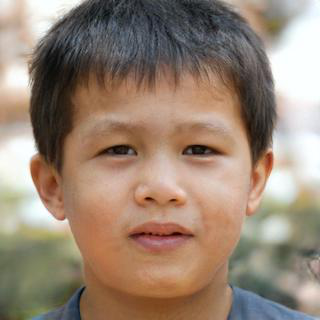} & \includegraphics[align=c,width=1.4cm]{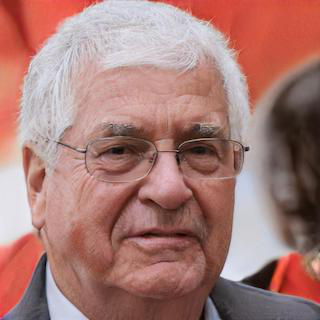} &  
    \includegraphics[align=c,width=1.4cm]{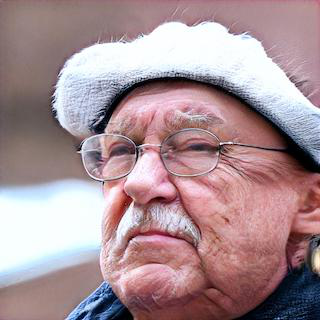} \\
     (c)&
     \includegraphics[align=c,width=1.4cm]{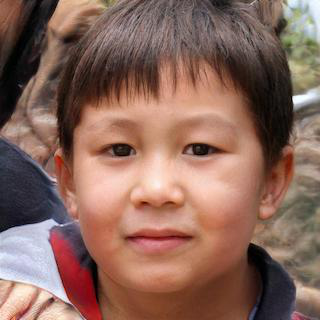} & \includegraphics[align=c,width=1.4cm]{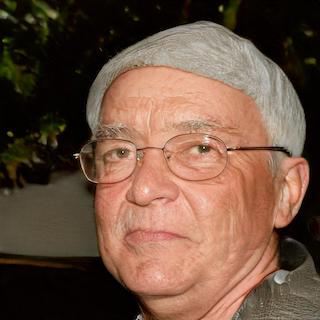} &
    \includegraphics[align=c,width=1.4cm]{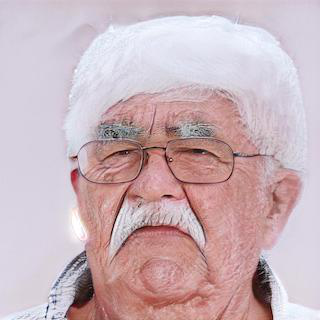} \\
     (d)&
     \includegraphics[align=c,width=1.4cm]{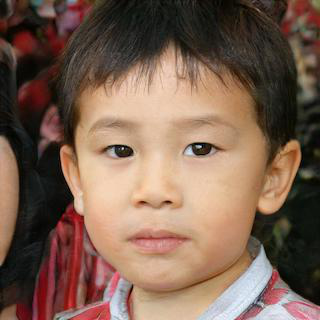} & \includegraphics[align=c,width=1.4cm]{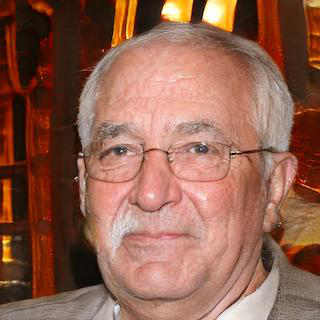} &
    \includegraphics[align=c,width=1.4cm]{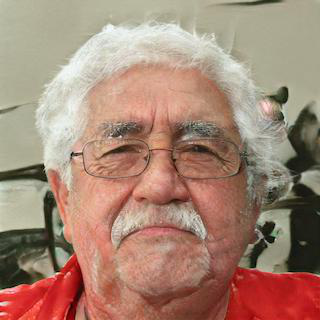} \\
     (e)&
     \includegraphics[align=c,width=1.4cm]{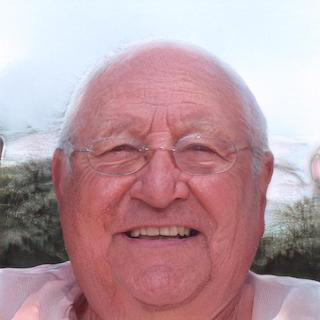} & \includegraphics[align=c,width=1.4cm]{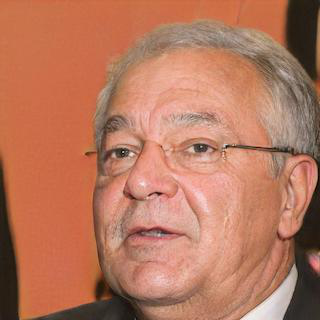} &
    \includegraphics[align=c,width=1.4cm]{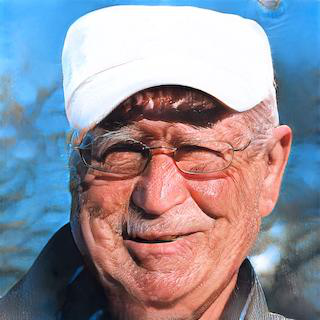} \\
     (f)&
     \includegraphics[align=c,width=1.4cm]{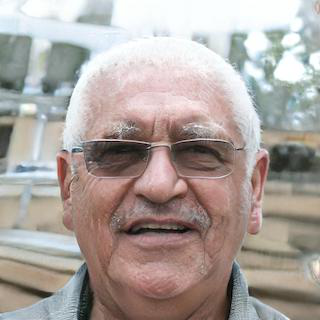} & \includegraphics[align=c,width=1.4cm]{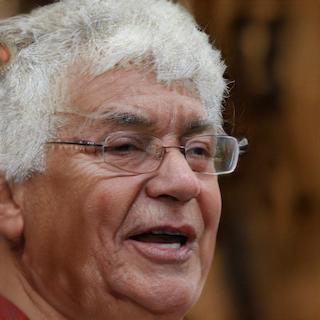} &
    \includegraphics[align=c,width=1.4cm]{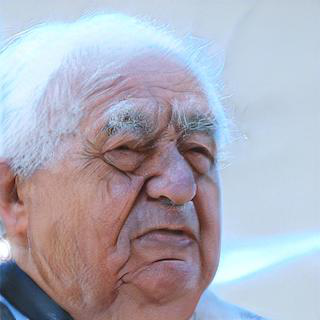} \\
    (g)&
     \includegraphics[align=c,width=1.4cm]{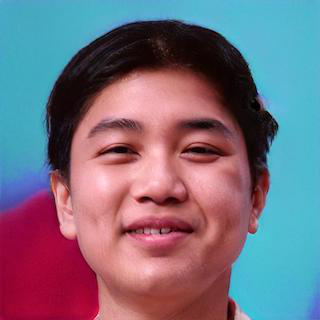} & \includegraphics[align=c,width=1.4cm]{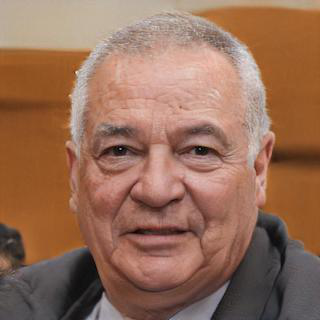}&
    \includegraphics[align=c,width=1.4cm]{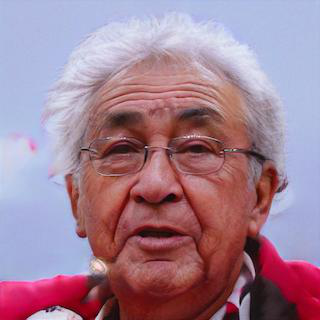} \\
     (h)&
     \includegraphics[align=c,width=1.4cm]{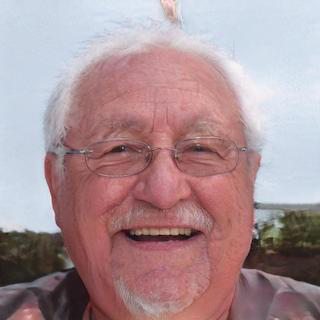} & \includegraphics[align=c,width=1.4cm]{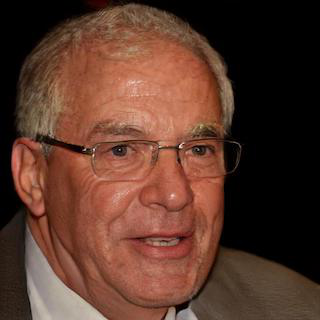} &
    \includegraphics[align=c,width=1.4cm]{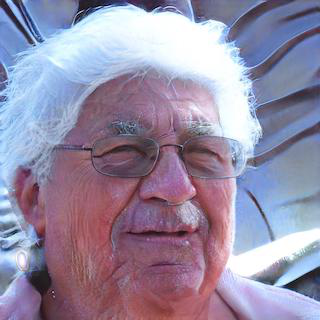} \\
     (i)&
     \includegraphics[align=c,width=1.4cm]{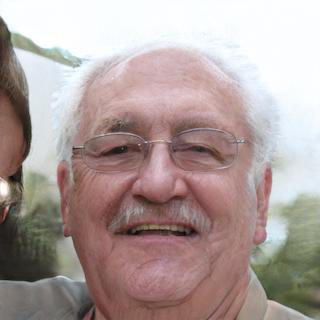} & \includegraphics[align=c,width=1.4cm]{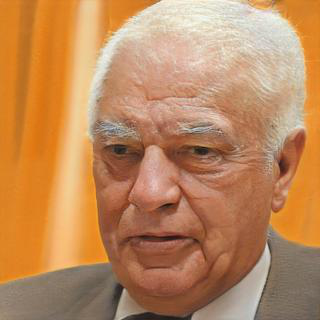} &
    \includegraphics[align=c,width=1.4cm]{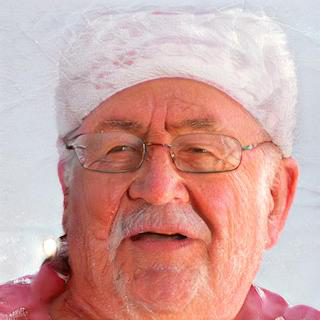} \\
    (j)&
     \includegraphics[align=c,width=1.4cm]{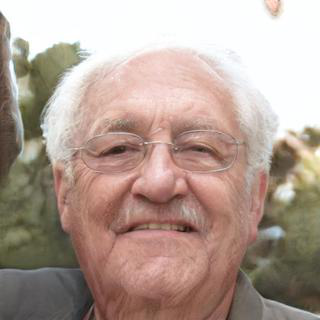} & \includegraphics[align=c,width=1.4cm]{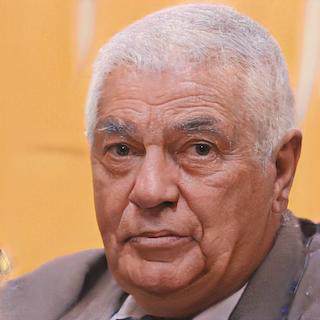} &
    \includegraphics[align=c,width=1.4cm]{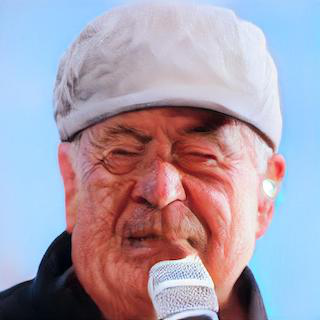} \\
    (k)&
     \includegraphics[align=c,width=1.4cm]{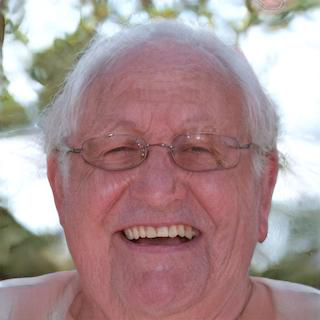} & \includegraphics[align=c,width=1.4cm]{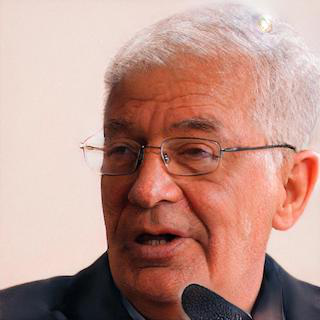} &
    \includegraphics[align=c,width=1.4cm]{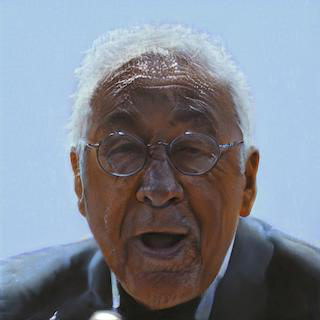} \\
\end{tabular}
\caption{Master face images with the highest MSC score, generated by (a) Random Search. (b) DE. (c) LSHADE-RSP. (d) IMODE. (e) CMA-ES. (f) NGOpt. (g) lq-CMA-ES. (h) ACM-ES. (i) LM-MA-ES. (j) LM-MA-ES + LQ-Filter. (k) LM-MA-ES + Success Predictor.}
\label{fig:gen_faces}
\end{figure}

In general, algorithms from the CMA-ES family perform better than other algorithms, such as the DE family. In particular, the original implementation of LM-MA-ES\cite{loshchilov2017limited} performs better than CMA-ES, as expected for a high-dimensional optimization problem. In fact, LM-MA-ES is the best performer among the original variants without additional assisting models.
It is worth mentioning that even though lq-CMA-ES achieved worse results than the original CMA-ES, as might be expected from its use of predicted fitness values instead of the actual ones, its training process is faster. 
In Fig.~\ref{fig:gen_faces}, we present the generated master face images with the highest MSC score for each of the face descriptors. 
\begin{figure*}
\centering
\begin{tabular}{@{~}l@{~}c@{~}c@{~}c@{~}c@{~}c@{~}c@{~}c@{~}c@{~}c@{~}}
          (a) &
    \includegraphics[align=c,width=1.4cm]{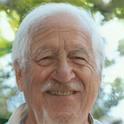} & \includegraphics[align=c,width=1.4cm]{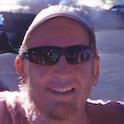} & \includegraphics[align=c,width=1.4cm]{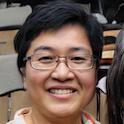} &      \includegraphics[align=c,width=1.4cm]{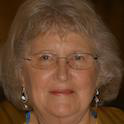} &   
    \includegraphics[align=c,width=1.4cm]{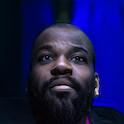} & \includegraphics[align=c,width=1.4cm]{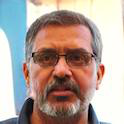} & \includegraphics[align=c,width=1.4cm]{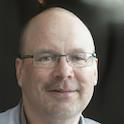} &      \includegraphics[align=c,width=1.4cm]{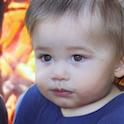} &   
    \includegraphics[align=c,width=1.4cm]{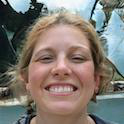} \\
      \small &$17.30\%$ & $9.84\%$ & $7.70\%$ & $6.39\%$ & $6.11\%$ & $5.29\%$ & $4.85\%$ & $3.45\%$ & $2.29\%$ \\
      (b) &
    \includegraphics[align=c,width=1.4cm]{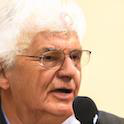} & \includegraphics[align=c,width=1.4cm]{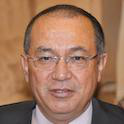} & \includegraphics[align=c,width=1.4cm]{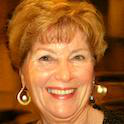} &      \includegraphics[align=c,width=1.4cm]{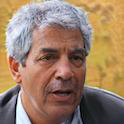} &   
    \includegraphics[align=c,width=1.4cm]{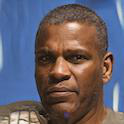} & \includegraphics[align=c,width=1.4cm]{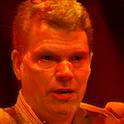} & \includegraphics[align=c,width=1.4cm]{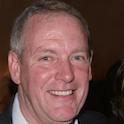} &     \includegraphics[align=c,width=1.4cm]{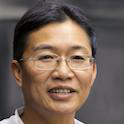} &   
    \includegraphics[align=c,width=1.4cm]{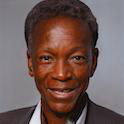} \\
      \small & $16.28\%$ & $5.70\%$ & $4.40\%$ & $3.56\%$ & $3.30\%$ & $2.90\%$ & $2.61\%$ & $1.91\%$ & $1.39\%$ \\
    (c) &
    \includegraphics[align=c,width=1.4cm]{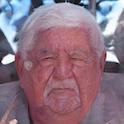} & \includegraphics[align=c,width=1.4cm]{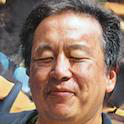} & \includegraphics[align=c,width=1.4cm]{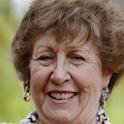} &  \includegraphics[align=c,width=1.4cm]{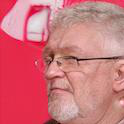} &
    \includegraphics[align=c,width=1.4cm]{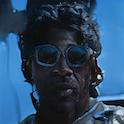} & \includegraphics[align=c,width=1.4cm]{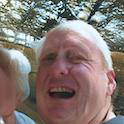} & \includegraphics[align=c,width=1.4cm]{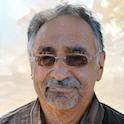} &  \includegraphics[align=c,width=1.4cm]{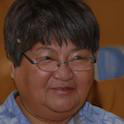} &
    \includegraphics[align=c,width=1.4cm]{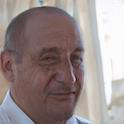} \\
    \small & $16.81\%$ & $5.50\%$ & $4.83\%$ & $3.42\%$ & $3.40\%$ & $3.09\%$ & $3.00\%$ & $1.60\%$ & $1.49\%$ \\
          (d) &    
\includegraphics[align=c,width=1.4cm]{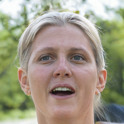} &
\includegraphics[align=c,width=1.4cm]{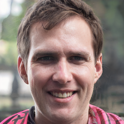} & 
\includegraphics[align=c,width=1.4cm]{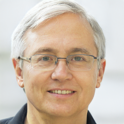} &
\includegraphics[align=c,width=1.4cm]{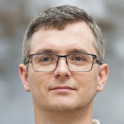} & 
    \includegraphics[align=c,width=1.4cm]{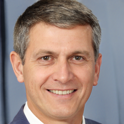} & 
    \includegraphics[align=c,width=1.4cm]{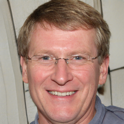} & 
    \includegraphics[align=c,width=1.4cm]{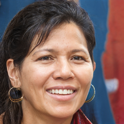}&
    \includegraphics[align=c,width=1.4cm]{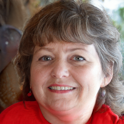} &
    \includegraphics[align=c,width=1.4cm]{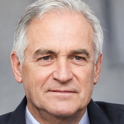} \\
    \small & $1.25\%$ & $1.04\%$ & $1.03\%$ & $0.97\%$ & $0.94\%$ & $0.87\%$ & $0.85\%$ & $0.8\%$ & $0.68\%$ \\
                (e) &    
\includegraphics[align=c,width=1.4cm]{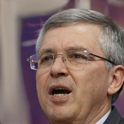} &
\includegraphics[align=c,width=1.4cm]{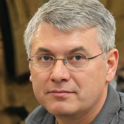} & 
\includegraphics[align=c,width=1.4cm]{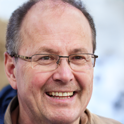} & 
\includegraphics[align=c,width=1.4cm]{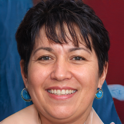} & 
    \includegraphics[align=c,width=1.4cm]{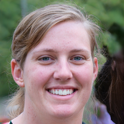} & 
    \includegraphics[align=c,width=1.4cm]{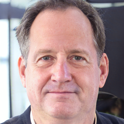} & 
    \includegraphics[align=c,width=1.4cm]{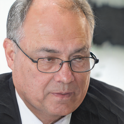}&
    \includegraphics[align=c,width=1.4cm]{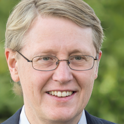} &
    \includegraphics[align=c,width=1.4cm]{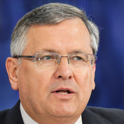} \\
    \small & $1.17\%$ & $0.98\%$ & $0.82\%$ & $0.82\%$ & $0.8\%$ & $0.78\%$ & $0.70\%$ & $0.64\%$ & $0.63\%$ \\
          (f) &    
\includegraphics[align=c,width=1.4cm]{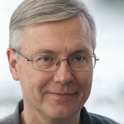} &
\includegraphics[align=c,width=1.4cm]{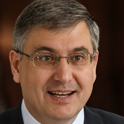} & 
\includegraphics[align=c,width=1.4cm]{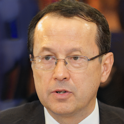} & 
\includegraphics[align=c,width=1.4cm]{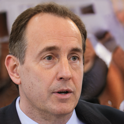} & 
    \includegraphics[align=c,width=1.4cm]{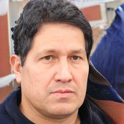} &
    \includegraphics[align=c,width=1.4cm]{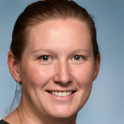} & 
    \includegraphics[align=c,width=1.4cm]{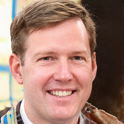}&
    \includegraphics[align=c,width=1.4cm]{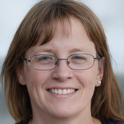} &
    \includegraphics[align=c,width=1.4cm]{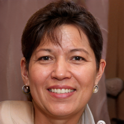} \\
    \small & $0.94\%$ & $0.78\%$ & $0.71\%$ & $0.70\%$ & $0.70\%$ & $0.70\%$ & $0.64\%$ & $0.64\%$ & $0.61\%$ \\
      (g) &
    \includegraphics[align=c,width=1.4cm]{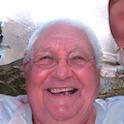} & \includegraphics[align=c,width=1.4cm]{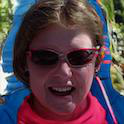} & \includegraphics[align=c,width=1.4cm]{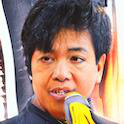} &      \includegraphics[align=c,width=1.4cm]{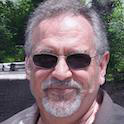} &   
    \includegraphics[align=c,width=1.4cm]{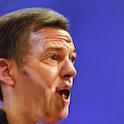} & \includegraphics[align=c,width=1.4cm]{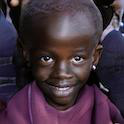} & \includegraphics[align=c,width=1.4cm]{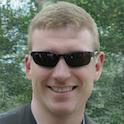} &      \includegraphics[align=c,width=1.4cm]{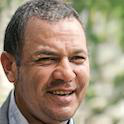} &   
    \includegraphics[align=c,width=1.4cm]{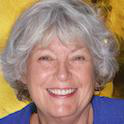} \\
      \small & $23.92\%$ & $7.07\%$ & $6.60\%$ & $6.01\%$ & $5.53\%$ & $5.34\%$ & $5.06\%$ & $2.80\%$ & $1.59\%$ \\
      (h) &
    \includegraphics[align=c,width=1.4cm]{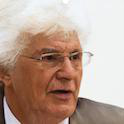} & \includegraphics[align=c,width=1.4cm]{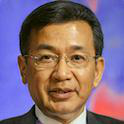} & \includegraphics[align=c,width=1.4cm]{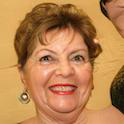} &      \includegraphics[align=c,width=1.4cm]{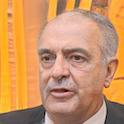} &   
    \includegraphics[align=c,width=1.4cm]{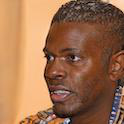} & \includegraphics[align=c,width=1.4cm]{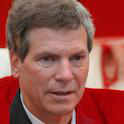} & \includegraphics[align=c,width=1.4cm]{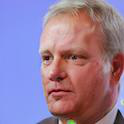} &     \includegraphics[align=c,width=1.4cm]{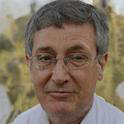} &   
    \includegraphics[align=c,width=1.4cm]{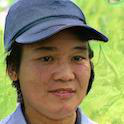} \\
      \small & $15.85\%$ & $5.05\%$ & $4.85\%$ & $4.00\%$ & $3.84\%$ & $3.67\%$ & $3.35\%$ & $1.61\%$ & $1.58\%$ \\
           (i) &
    \includegraphics[align=c,width=1.4cm]{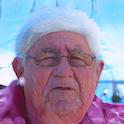} & \includegraphics[align=c,width=1.4cm]{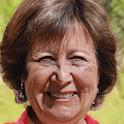} & \includegraphics[align=c,width=1.4cm]{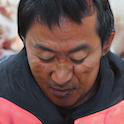} &  \includegraphics[align=c,width=1.4cm]{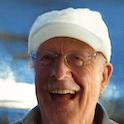} &
    \includegraphics[align=c,width=1.4cm]{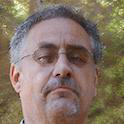} & \includegraphics[align=c,width=1.4cm]{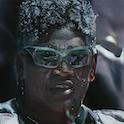} & \includegraphics[align=c,width=1.4cm]{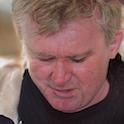} &  \includegraphics[align=c,width=1.4cm]{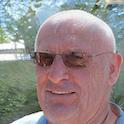} &
    \includegraphics[align=c,width=1.4cm]{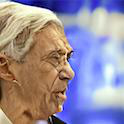} \\
    \small & $17.17\%$ & $4.75\%$ & $4.63\%$ & $4.45\%$ & $3.65\%$ & $3.23\%$ & $2.20\%$ & $2.10\%$ & $1.97\%$ \\
            (j) &    
\includegraphics[align=c,width=1.4cm]{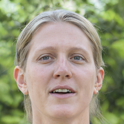} & 
\includegraphics[align=c,width=1.4cm]{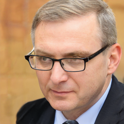} & 
\includegraphics[align=c,width=1.4cm]{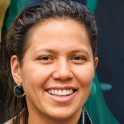} &
\includegraphics[align=c,width=1.4cm]{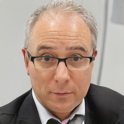} & 
    \includegraphics[align=c,width=1.4cm]{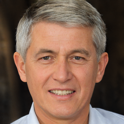} & 
    \includegraphics[align=c,width=1.4cm]{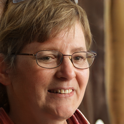} &
    \includegraphics[align=c,width=1.4cm]{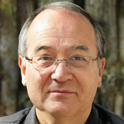}&
    \includegraphics[align=c,width=1.4cm]{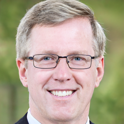} &
    \includegraphics[align=c,width=1.4cm]{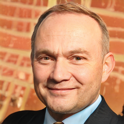} \\
    \small & $1.2\%$ & $1.08\%$ & $0.97\%$ & $0.96\%$ & $0.96\%$ & $0.92\%$ & $0.89\%$ & $0.87\%$ & $0.78\%$ \\
          (k) &    
\includegraphics[align=c,width=1.4cm]{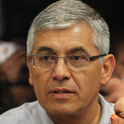} &
\includegraphics[align=c,width=1.4cm]{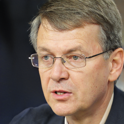} & 
\includegraphics[align=c,width=1.4cm]{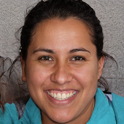} & 
\includegraphics[align=c,width=1.4cm]{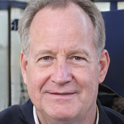} & 
    \includegraphics[align=c,width=1.4cm]{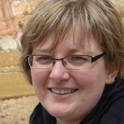} &
    \includegraphics[align=c,width=1.4cm]{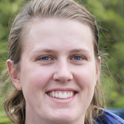} & 
    \includegraphics[align=c,width=1.4cm]{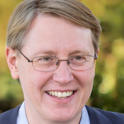}&
    \includegraphics[align=c,width=1.4cm]{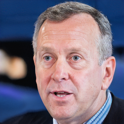} &
    \includegraphics[align=c,width=1.4cm]{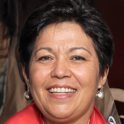} \\
    \small & $1.13\%$ & $0.99\%$ & $0.82\%$ & $0.70\%$ & $0.70\%$ & $0.66\%$ & $0.66\%$ & $0.64\%$ & $0.63\%$ \\
          (l) &    
\includegraphics[align=c,width=1.4cm]{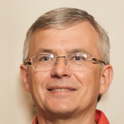} &
\includegraphics[align=c,width=1.4cm]{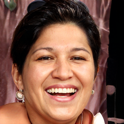} & 
\includegraphics[align=c,width=1.4cm]{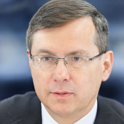} & 
\includegraphics[align=c,width=1.4cm]{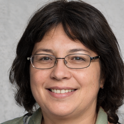} & 
    \includegraphics[align=c,width=1.4cm]{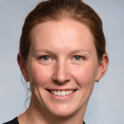} &
    \includegraphics[align=c,width=1.4cm]{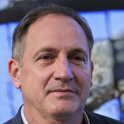} & 
    \includegraphics[align=c,width=1.4cm]{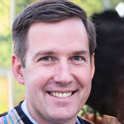}&
    \includegraphics[align=c,width=1.4cm]{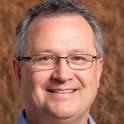} &
    \includegraphics[align=c,width=1.4cm]{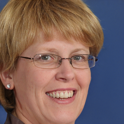} \\
    \small & $1.2\%$ & $0.77\%$ & $0.71\%$ & $0.66\%$ & $0.66\%$ & $0.64\%$ & $0.63\%$ & $0.59\%$ & $0.54\%$ \\
\end{tabular}
\caption{Set of nine master face images generated with each of the Coverage Search methods on LFW: greedy-Coverage Search methods (a-f) LM-MA-ES, (g-l) LM-MA-ES+Success Predictor, embedded using either (a,g) Dlib, (b,h) FaceNet, (c,i) SphereFace, (d,j) MagFace, (e,k) ElasticFace-Arc or (f,l) the ElasticFace-Cos face descriptors. Below each image, its MSC score is listed.}\label{fig:cover_image}
\end{figure*}

\subsection{Results}
We run Alg.~\ref{alg:greedy} to find a minimal number of images that match the largest number of faces in the entire dataset $D$ of LFW, where $|D|=5,749$. 

As a baseline, we clustered the embedding faces to nine clusters. Based on these clusters, we split the dataset into nine disjoint datasets. We run the single image generation method (Fig~\ref{fig:alg_flow}) on each dataset. Our $greedy$ method (Alg.~\ref{alg:greedy}) is also run for nine iterations in order to generate a number of faces equal to the number of clusters. We run it twice, once with the LM-MA-ES optimization method, and once with our LM-MA-ES+Success Predictor.

\begin{table*}
\caption{Percentage of the LFW dataset covered by nine generated images}
\label{table:ds_cover}
\begin{center}
\begin{tabular}{lcccccc}
\toprule
 & Dlib & FaceNet & SphereFace & {MagFace} & {ElasticFace-Arc} &{ElasticFace-Cos} \\
\midrule
\makecell[l]{Coverage Search LM-MA-ES\\on clustered data}& $51.43\%$ &  $40.17\%$ & $39.78\%$ &$5.29\%$ &$4.85\%$ &$3.91\%$\\ 
\hline
\makecell[l]{$greedy$-Coverage Search\\ LM-MA-ES}& $63.22\%$ &  $42.08\%$ & $43.14\%$ &$8.44\%$ &$\mathbf{7.32}\%$ &$\mathbf{6.42}\%$\\ 
\hline
\makecell[l]{$greedy$-Coverage Search\\ LM-MA-ES+Success Pred.}& $\mathbf{63.92}\%$&  $\mathbf{43.82}\%$ & $\mathbf{44.15}\%$ &$\mathbf{8.63}\%$ &$6.92\%$ &$6.40\%$\\ 
\bottomrule
\end{tabular}
\end{center}
\vspace{-4mm}
\end{table*}
Table~\ref{table:ds_cover} lists the results. Evidently, the greedy method outperforms the per-cluster solution. 
Applying the Success Predictor slightly improves results for {four out of the six} face descriptors. The coverage results for Dlib is larger than the other face descriptors. The higher embedding dimension of the other face descriptors allows better separability between faces, which in turn makes the coverage task more difficult. {The coverage results for the newer models, MagFace and ElasticFace, are considerably lower than the other models, but still posses a considerable weakness to these systems, that exists at a low FAR of $\sim0.001$. } Fig.~\ref{fig:cover_image} presents nine master face images for each of the face descriptors, generated by $greedy$-Coverage Search algorithms, together with their MSC scores.
{Specifically, the master faces are sorted from left to right in a descending order with respect to their MSC score. The left most master face achieves the highest MSC score, i.e. this master face succeeds in being incorrectly authorized by the largest number of identities in the dataset. For any other master face, we do not make double counting, i.e. we do not take into consideration identities that are already covered by better master faces in terms of dataset coverage. As a result, the sum of MSC scores of all nine master faces in a given row in Figure \ref{fig:cover_image} is equal to the corresponding result that is reported in Table \ref{table:ds_cover}}.

\begin{figure*}
\centering
\begin{tabular}{@{~}l@{~}c@{~}c@{~}c@{~}c@{~}c@{~}c@{~}c@{~}c@{~}c@{~}}
    (a) &
\includegraphics[align=c,width=1.4cm]{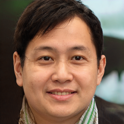} &
\includegraphics[align=c,width=1.4cm]{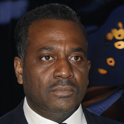} &
\includegraphics[align=c,width=1.4cm]{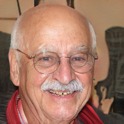} &
\includegraphics[align=c,width=1.4cm]{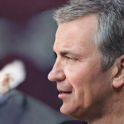} &
\includegraphics[align=c,width=1.4cm]{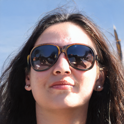} &
\includegraphics[align=c,width=1.4cm]{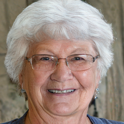} &
\includegraphics[align=c,width=1.4cm]{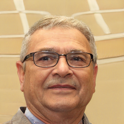} &
\includegraphics[align=c,width=1.4cm]{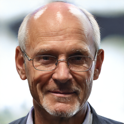} &
\includegraphics[align=c,width=1.4cm]{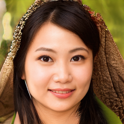} \\
\small &$34.08\%$ &$13.96\%$ &$12.86\%$ &$5.72\%$ &$5.53\%$ &$3.15\%$ &$2.12\%$ &$1.16\%$ &$1.03\%$ \\ 
    (b) &
\includegraphics[align=c,width=1.4cm]{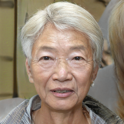} &
\includegraphics[align=c,width=1.4cm]{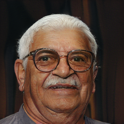} &
\includegraphics[align=c,width=1.4cm]{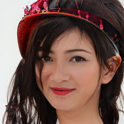} &
\includegraphics[align=c,width=1.4cm]{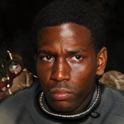} &
\includegraphics[align=c,width=1.4cm]{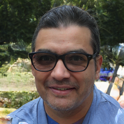} &
\includegraphics[align=c,width=1.4cm]{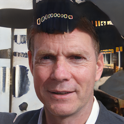} &
\includegraphics[align=c,width=1.4cm]{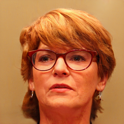} &
\includegraphics[align=c,width=1.4cm]{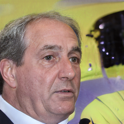} &
\includegraphics[align=c,width=1.4cm]{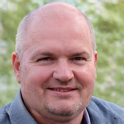} \\
\small &$21.36\%$ &$13.15\%$ &$10.90\%$ &$9.49\%$ &$4.17\%$ &$3.59\%$ &$1.99\%$ &$1.54\%$ &$1.09\%$ \\ 
           (c) &
\includegraphics[align=c,width=1.4cm]{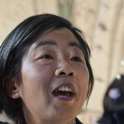} &
\includegraphics[align=c,width=1.4cm]{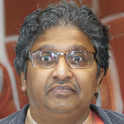} &
\includegraphics[align=c,width=1.4cm]{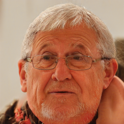} &
\includegraphics[align=c,width=1.4cm]{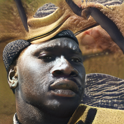} &
\includegraphics[align=c,width=1.4cm]{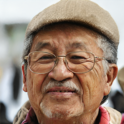} &
\includegraphics[align=c,width=1.4cm]{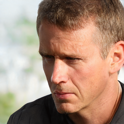} &
\includegraphics[align=c,width=1.4cm]{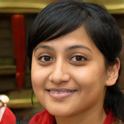} &
\includegraphics[align=c,width=1.4cm]{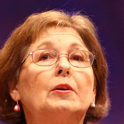} &
\includegraphics[align=c,width=1.4cm]{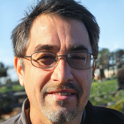} \\
\small &$25.59\%$ &$14.18\%$ &$9.88\%$ &$8.85\%$ &$2.50\%$ &$2.44\%$ &$2.37\%$ &$1.92\%$ &$1.60\%$ \\ 

           (d) &
\includegraphics[align=c,width=1.4cm]{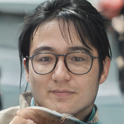} &
\includegraphics[align=c,width=1.4cm]{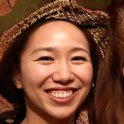} &
\includegraphics[align=c,width=1.4cm]{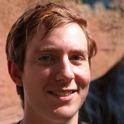} &
\includegraphics[align=c,width=1.4cm]{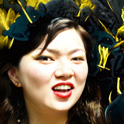} &
\includegraphics[align=c,width=1.4cm]{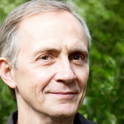} &
\includegraphics[align=c,width=1.4cm]{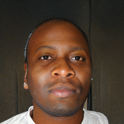} &
\includegraphics[align=c,width=1.4cm]{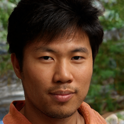} &
\includegraphics[align=c,width=1.4cm]{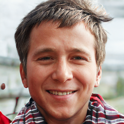} &
\includegraphics[align=c,width=1.4cm]{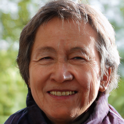} \\
\small &$2.82\%$ &$2.37\%$ &$2.05\%$ &$1.86\%$ &$1.67\%$ &$1.60\%$ &$1.35\%$ &$1.28\%$ &$0.96\%$ \\ 

           (e) &
\includegraphics[align=c,width=1.4cm]{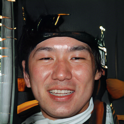} &
\includegraphics[align=c,width=1.4cm]{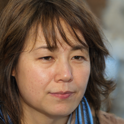} &
\includegraphics[align=c,width=1.4cm]{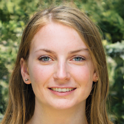} &
\includegraphics[align=c,width=1.4cm]{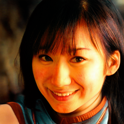} &
\includegraphics[align=c,width=1.4cm]{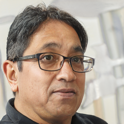} &
\includegraphics[align=c,width=1.4cm]{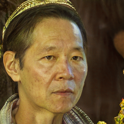} &
\includegraphics[align=c,width=1.4cm]{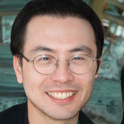} &
\includegraphics[align=c,width=1.4cm]{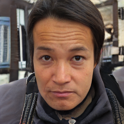} &
\includegraphics[align=c,width=1.4cm]{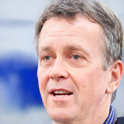} \\
\small &$1.16\%$ &$1.03\%$ &$0.77\%$ &$0.64\%$ &$0.58\%$ &$0.58\%$ &$0.51\%$ &$0.45\%$ &$0.32\%$ \\ 

           (f) &
\includegraphics[align=c,width=1.4cm]{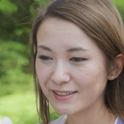} &
\includegraphics[align=c,width=1.4cm]{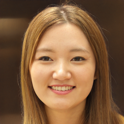} &
\includegraphics[align=c,width=1.4cm]{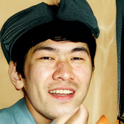} &
\includegraphics[align=c,width=1.4cm]{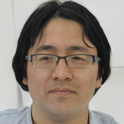} &
\includegraphics[align=c,width=1.4cm]{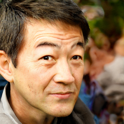} &
\includegraphics[align=c,width=1.4cm]{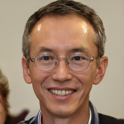} &
\includegraphics[align=c,width=1.4cm]{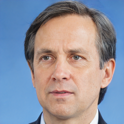} &
\includegraphics[align=c,width=1.4cm]{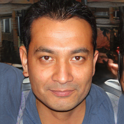} &
\includegraphics[align=c,width=1.4cm]{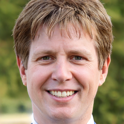} \\
\small &$1.54\%$ &$1.09\%$ &$1.03\%$ &$0.96\%$ &$0.77\%$ &$0.45\%$ &$0.39\%$ &$0.39\%$ &$0.19\%$ \\ 
       (g) &
\includegraphics[align=c,width=1.4cm]{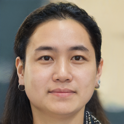} &
\includegraphics[align=c,width=1.4cm]{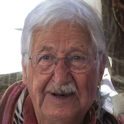} &
\includegraphics[align=c,width=1.4cm]{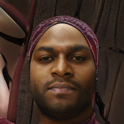} &
\includegraphics[align=c,width=1.4cm]{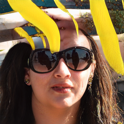} &
\includegraphics[align=c,width=1.4cm]{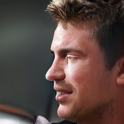} &
\includegraphics[align=c,width=1.4cm]{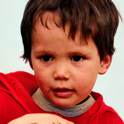} &
\includegraphics[align=c,width=1.4cm]{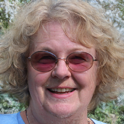} &
\includegraphics[align=c,width=1.4cm]{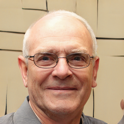} &
\includegraphics[align=c,width=1.4cm]{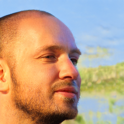} \\
\small &$31.45\%$ &$16.79\%$ &$14.15\%$ &$7.20\%$ &$6.50\%$ &$3.79\%$ &$2.70\%$ &$1.29\%$ &$0.90\%$ \\ 

           (h) &
\includegraphics[align=c,width=1.4cm]{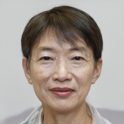} &
\includegraphics[align=c,width=1.4cm]{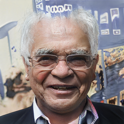} &
\includegraphics[align=c,width=1.4cm]{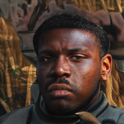} &
\includegraphics[align=c,width=1.4cm]{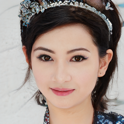} &
\includegraphics[align=c,width=1.4cm]{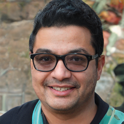} &
\includegraphics[align=c,width=1.4cm]{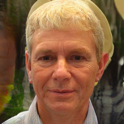} &
\includegraphics[align=c,width=1.4cm]{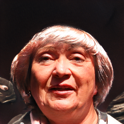} &
\includegraphics[align=c,width=1.4cm]{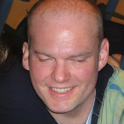} &
\includegraphics[align=c,width=1.4cm]{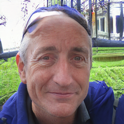} \\
\small &$21.55\%$ &$15.71\%$ &$9.81\%$ &$8.47\%$ &$5.13\%$ &$3.66\%$ &$2.89\%$ &$1.67\%$ &$1.16\%$ \\ 

           (i) &
\includegraphics[align=c,width=1.4cm]{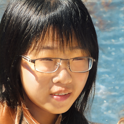} &
\includegraphics[align=c,width=1.4cm]{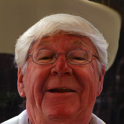} &
\includegraphics[align=c,width=1.4cm]{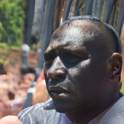} &
\includegraphics[align=c,width=1.4cm]{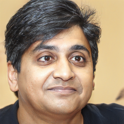} &
\includegraphics[align=c,width=1.4cm]{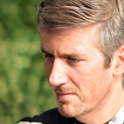} &
\includegraphics[align=c,width=1.4cm]{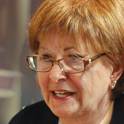} &
\includegraphics[align=c,width=1.4cm]{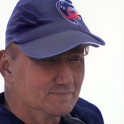} &
\includegraphics[align=c,width=1.4cm]{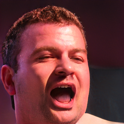} &
\includegraphics[align=c,width=1.4cm]{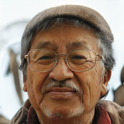} \\
\small &$26.17\%$ &$14.11\%$ &$11.55\%$ &$8.72\%$ &$3.21\%$ &$1.99\%$ &$1.80\%$ &$1.35\%$ &$1.09\%$ \\ 

           (j) &
\includegraphics[align=c,width=1.4cm]{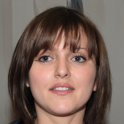} &
\includegraphics[align=c,width=1.4cm]{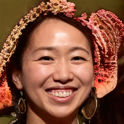} &
\includegraphics[align=c,width=1.4cm]{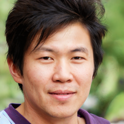} &
\includegraphics[align=c,width=1.4cm]{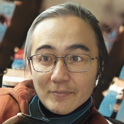} &
\includegraphics[align=c,width=1.4cm]{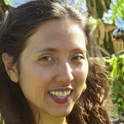} &
\includegraphics[align=c,width=1.4cm]{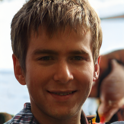} &
\includegraphics[align=c,width=1.4cm]{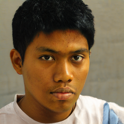} &
\includegraphics[align=c,width=1.4cm]{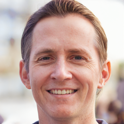} &
\includegraphics[align=c,width=1.4cm]{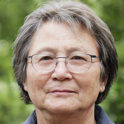} \\
\small &$2.63\%$ &$2.18\%$ &$1.92\%$ &$1.73\%$ &$1.54\%$ &$1.48\%$ &$1.28\%$ &$1.03\%$ &$0.90\%$ \\

           (k) &
\includegraphics[align=c,width=1.4cm]{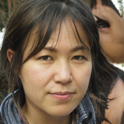} &
\includegraphics[align=c,width=1.4cm]{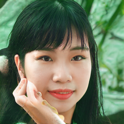} &
\includegraphics[align=c,width=1.4cm]{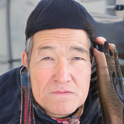} &
\includegraphics[align=c,width=1.4cm]{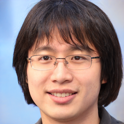} &
\includegraphics[align=c,width=1.4cm]{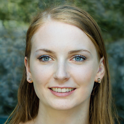} &
\includegraphics[align=c,width=1.4cm]{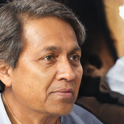} &
\includegraphics[align=c,width=1.4cm]{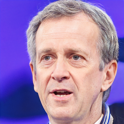} &
\includegraphics[align=c,width=1.4cm]{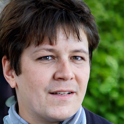} &
\includegraphics[align=c,width=1.4cm]{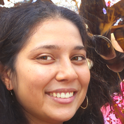} \\
\small &$1.28\%$ &$0.96\%$ &$0.77\%$ &$0.64\%$ &$0.64\%$ &$0.58\%$ &$0.51\%$ &$0.45\%$ &$0.39\%$ \\ 

           (l) &
\includegraphics[align=c,width=1.4cm]{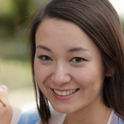} &
\includegraphics[align=c,width=1.4cm]{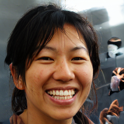} &
\includegraphics[align=c,width=1.4cm]{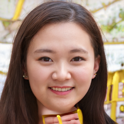} &
\includegraphics[align=c,width=1.4cm]{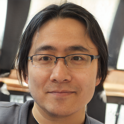} &
\includegraphics[align=c,width=1.4cm]{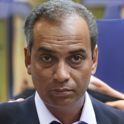} &
\includegraphics[align=c,width=1.4cm]{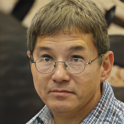} &
\includegraphics[align=c,width=1.4cm]{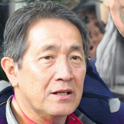} &
\includegraphics[align=c,width=1.4cm]{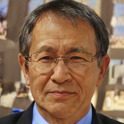} &
\includegraphics[align=c,width=1.4cm]{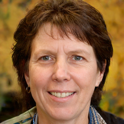} \\
\small &$1.35\%$ &$1.16\%$ &$0.83\%$ &$0.77\%$ &$0.77\%$ &$0.71\%$ &$0.51\%$ &$0.45\%$ &$0.39\%$ \\

 \end{tabular}
\caption{Set of nine master face images generated with each of the Coverage Search methods on the RFW subset: greedy-Coverage Search methods (a-f) LM-MA-ES, (g-l) LM-MA-ES+Success Predictor, embedded using either (a,g) Dlib, (b,h) FaceNet, (c,i) SphereFace, (d,j) MagFace, (e,k) ElasticFace-Arc or (f,l) the ElasticFace-Cos face descriptors. Below each image, its test MSC score is listed.}\label{fig:cover_image_rfw}
\end{figure*}

{For the purpose of evaluating our method in a more ethnically-diverse setting, we run the same experiments  for generating nine master faces, which cover the maximal portion of the Racial Faces in the Wild (RFW) dataset~\cite{Wang_2019_ICCV}. RFW consists of four balanced ethnic subsets (African, Caucasian, Asian and Indian) of face images that were collected from MS-Celeb-1M~\cite{guo2016ms} for evaluating the racial bias of face recognition models. Following Wang et al.~\cite{wang2021meta}, we sub-sample RFW according to the distribution of these four ethnic groups in the real world population. For evaluating the generalization of our model, we split the obtained subset to training and testing subsets of sizes $|D_{train}|=6223$ and $|D_{test}|=1559$, respectively. 

{To establish an approximation of the performance in the case in which the output need not be inverted into an image, we first performed the attack in the embedding space instead of the image space. In this case, using nine clusters in the RFW embedding space, the obtained coverage ranges from a maximum of $99.24\%$ for Dlib to a minimum of $22.96\%$ for ElasticFace-Arc.}

Table \ref{table:ds_cover_rfw} lists the mean set coverage results obtained for each of the mentioned experiments. Evidently, our suggested LM-MA-ES+Success Predictor succeeds in covering a larger portion of the training subset than the baselines for five out of six face recognition models. More importantly, our suggested LM-MA-ES+Success Predictor generated master faces that generalize better than the baselines and they cover a larger portion of the testing subset also for five out of six face recognition models. Figure \ref{fig:cover_image_rfw} presents the nine master face images that were generated for each face descriptor by each method on the subset of RFW dataset, whose ethnic distribution is similar to the real-world distribution.
}
\begin{table*}
\centering
{
\caption{Percentage of the RFW subset covered by nine generated images}
\label{table:ds_cover_rfw}
\begin{tabular}{lcccccccccccc} 
\toprule
\multirow{2}{*}{}                                         & \multicolumn{2}{c}{Dlib}                             & \multicolumn{2}{c}{FaceNet}                          & \multicolumn{2}{c}{SphereFace}                       & \multicolumn{2}{c}{MagFace}                          & \multicolumn{2}{c}{ElasticFace-Arc}                  & \multicolumn{2}{c}{ElasticFace-Cos}                   \\ 
\cline{2-13}
& \multicolumn{1}{c}{Train} & \multicolumn{1}{c}{Test} & \multicolumn{1}{c}{Train} & \multicolumn{1}{c}{Test} & \multicolumn{1}{c}{Train} & \multicolumn{1}{c}{Test} & \multicolumn{1}{c}{Train} & \multicolumn{1}{c}{Test} & \multicolumn{1}{c}{Train} & \multicolumn{1}{c}{Test} & \multicolumn{1}{c}{Train} & \multicolumn{1}{c}{Test}  \\ 
\midrule
\begin{tabular}[c]{@{}l@{}}Coverage Search LM-MA-ES\\on clustered data\end{tabular} &$83.02$ &$81.61$ &$67.56$ &$65.81$ &$62.49$ &$59.46$ &$17.93$ &$12.70$ &$10.40$ &$5.77$ &$10.69$ &$6.54$  \\ 
\midrule
\begin{tabular}[c]{@{}l@{}}greedy-Coverage Search\\LM-MA-ES\end{tabular}  &$80.80$ &$79.61$ &$72.07$ &$67.29$ &$72.30$ &$69.34$ &$22.10$ &$\mathbf{15.97}$ &$\mathbf{14.38}$ &$6.03$ &$14.75$ &$6.80$  \\ 
\midrule
\begin{tabular}[c]{@{}l@{}}greedy-Coverage Search\\LM-MA-ES+Success Pred\end{tabular}  &$\mathbf{84.16}$ &$\mathbf{84.76}$ &$\mathbf{72.60}$ &$\mathbf{70.04}$ &$\mathbf{73.31}$ &$\mathbf{69.98}$ &$\mathbf{22.58}$ &$14.69$ &$13.24$ &$\mathbf{6.22}$ &$\mathbf{14.93}$ &$\mathbf{6.93}$  \\ 
\bottomrule
\end{tabular}}
\end{table*}

\section{Experiments - 3D Scenario}
\label{subsec:3d_scenario}
In order to further demonstrate the effectiveness of our method, we evaluated it in the scenario of 3D face recognition, specifically, with the recognition systems of Kim et al. \cite{kim2017deep} and FR3DNet \cite{FaceRec3D}.

Our method is evaluated on the Texas 3D Face Recognition Database \cite{gupta2010texas,gupta2010anthropometric}. Texas 3D is a dataset of 1,149 high-resolution and aligned pairs of color and range images. The 3D models of this dataset were acquired using a stereo imaging system and were pre-processed into a useful form for 3D face recognition. Texas 3D consists of neutral and expressive face images of 116 adult subjects, both male and female, within the age range of 22-75 years and from five ethnic groups (Caucasians, Africans, Asians, East Indians and Hispanics).

As the recognition threshold ($\theta$) in our 3D experiments, we employ the Equal Error Rate (ERR) trade-off point, i.e. the classification threshold for which the FAR is approximately equal to the FRR. For Kim et al. (FR3DNet), the EER on the (unfiltered) Texas3D dataset is 0.036 (0.009).

For 3D face experiments, our optimizer (Eq. \ref{eq:z_opt}) requires the use of a 3D face generator as $\G$. While 2D face generation has been extensively researched and there are many successful publicly available generative models, there are far fewer options for 3D faces. In this work, we suggest modelling the 3D generative model using a 2D face generator followed by a 3D reconstruction model.

In particular, we use StyleGAN2 \cite{Karras2019stylegan2}, which was trained on 256x256 images of FFHQ \cite{styleGan}, as the 2D face generator. We then employ MTCNN \cite{mtcnn} for estimating the key facial landmarks of the synthesized 2D face image. These are passed to the reconstruction model of Deng et al. \cite{deng2019accurate} to obtain the corresponding 3D structure of the given synthesized 2D face. 

Deng et al. \cite{deng2019accurate} generate high-quality 3D face reconstructions by feeding Basel \cite{paysan20093d}, a 3D morphable face model, with parameters that were estimated using a deep convolutional network given the 2D input image. The PyTorch implementation of \cite{deng2019accurate} is reported to achieve the second best result in the NoW challenge \cite{sanyal2019learning}. The depth image and normals of the 3D reconstruction are computed using the PyTorch3D renderer \cite{ravi2020pytorch3d}. The estimated landmark of the nose is used to center the 3D face image, as required by FR3DNet.

\subsection{Dataset Coverage}\label{subsubec:3d_coverage}
We consider the dataset $D$, a subset of the Texas 3D dataset, containing a single 3D scan for each unique subject out of the 116 subjects, i.e., $|D|=116$. 

Firstly, we evaluate the clustering results on $D$ in the embedding space, similarly to Sec. \ref{subsec:cent_inv}. Since both Kim et al. \cite{kim2017deep} and FR3DNet \cite{FaceRec3D} use the cosine distance as the metric for comparing embedding vectors, we apply Spherical K-Means \cite{dhillon2001concept,kim2020improving}, an adaptation of the traditional K-Means algorithm for the cosine distance metric. 

With nine clusters, 91.38\% and 64.66\% of the embedded dataset are covered for Kim et al. and FR3DNet, respectively. These results provide an estimate of the achievable coverage of the embedded dataset, in an ideal case where a dictionary attack could be made on the embedding space instead of real-world 3D face images. As we demonstrated in \ref{subsec:cent_inv}, inverting such centroids from the embedding space to the image space is a difficult task, even for 2D images. Therefore, we run Alg. \ref{alg:greedy}, in order to find the maximal number of 3D images within the dataset $D$ that can be covered by at most nine 3D master faces, where $|D|=116$. 

We repeat the experiments for the 3D scenario. Specifically, the dataset is split into nine disjoint clusters according to the nine centroids computed by Spherical K-Means, and a single 3D master face is generated for each disjoint cluster. In addition, we run our $greedy$ method with the original variant of LM-MA-ES and with our suggested LM-MA-ES+Success Predictor.

The architecture and the hyperparameters used for the Success Predictor are identical to those employed in the 2D case, i.e., a feed-forward neural network consisting of three fully-connected layers, where the first two hidden layers and the output layer are followed by the ELU \cite{clevert2015fast} activation function and a Sigmoid, respectively.  

Table \ref{table:ds_3d_cover} lists the mean set coverage results obtained for each of the above mentioned experiments. Evidently, results for the 3D scenario are consistent with the results obtained for the 2D scenario, and are even more in a favor of our suggested LM-MA-ES+Success Predictor over the baselines. In general, the two runs of the $greedy$-coverage approach succeed in covering a large portion of the 3D dataset (40\%-50\%). Specifically, they outperform the baseline for optimizing master faces master face on the disjoint clusters.

\begin{figure*}
\centering
\begin{tabular}{@{~}l@{~}c@{~}c@{~}c@{~}c@{~}c@{~}c@{~}c@{~}c@{~}c@{~}}
   (a) &
    \includegraphics[align=c,width=1.8cm]{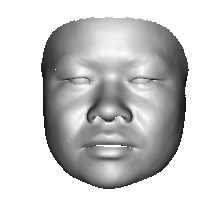} &    
    \includegraphics[align=c,width=1.8cm]{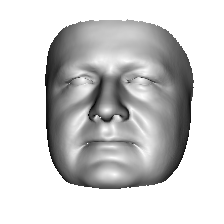} &    
    \includegraphics[align=c,width=1.8cm]{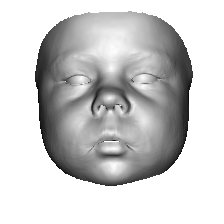} &    
    \includegraphics[align=c,width=1.8cm]{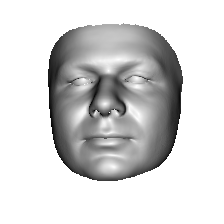} &    
    \includegraphics[align=c,width=1.8cm]{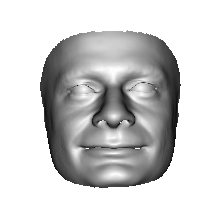} &    
    \includegraphics[align=c,width=1.8cm]{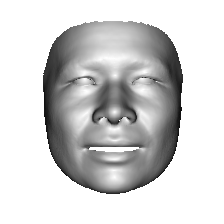} &    
    \includegraphics[align=c,width=1.8cm]{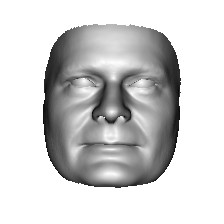} &    
    \\
    \small &$15.52\%$ & $7.76\%$ & $4.31\%$ & $4.31\%$ & $4.31\%$ & $0.86\%$ & $0.86\%$  \\
    (b) &
    \includegraphics[align=c,width=1.8cm]{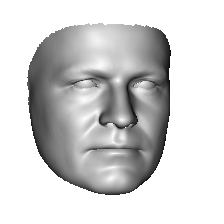} &    
    \includegraphics[align=c,width=1.8cm]{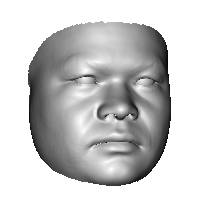} &   
    \includegraphics[align=c,width=1.8cm]{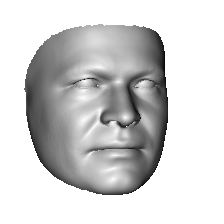} &   
    \includegraphics[align=c,width=1.8cm]{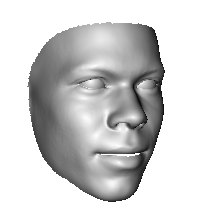} &   
    \includegraphics[align=c,width=1.8cm]{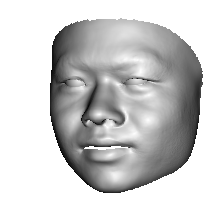} &   
    \includegraphics[align=c,width=1.8cm]{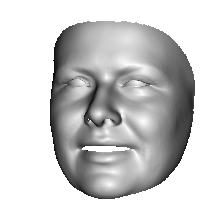} &   
    \includegraphics[align=c,width=1.8cm]{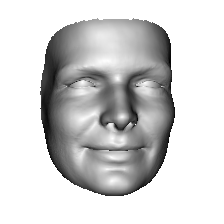} &   
    \\
    \small &$12.07\%$ & $11.21\%$ & $5.17\%$ & $5.17\%$ & $0.86\%$ & $0.86\%$ & $0.86\%$  \\
   (c) &
    \includegraphics[align=c,width=1.8cm]{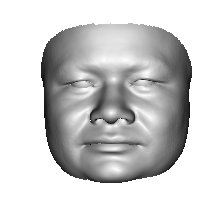} &     \includegraphics[align=c,width=1.8cm]{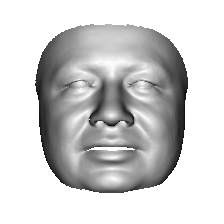} &
    \includegraphics[align=c,width=1.8cm]{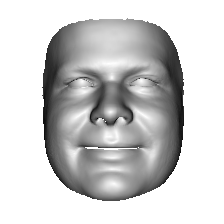} &
    \includegraphics[align=c,width=1.8cm]{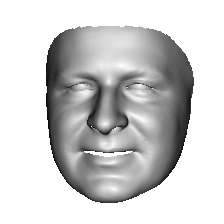} &
    \includegraphics[align=c,width=1.8cm]{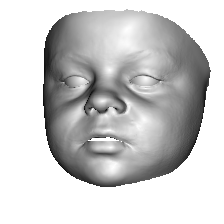} &
    \includegraphics[align=c,width=1.8cm]{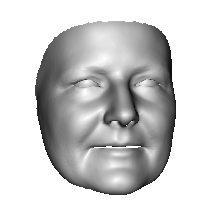} &
    \includegraphics[align=c,width=1.8cm]{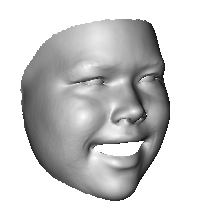} &
    \includegraphics[align=c,width=1.8cm]{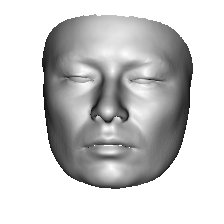} &
    \\
    \small &$26.72\%$ & $7.76\%$ & $1.72\%$ & $1.72\%$ & $0.86\%$ & $0.86\%$ & $0.86\%$ & $0.86\%$ & \\
     (d) &
    \includegraphics[align=c,width=1.8cm]{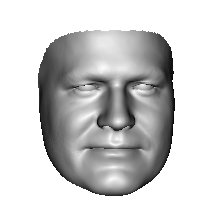} &     \includegraphics[align=c,width=1.8cm]{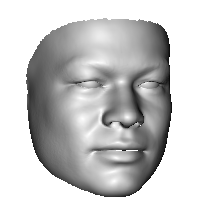} &
    \includegraphics[align=c,width=1.8cm]{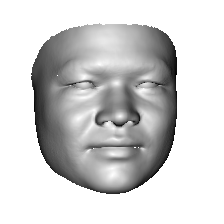} &
    \includegraphics[align=c,width=1.8cm]{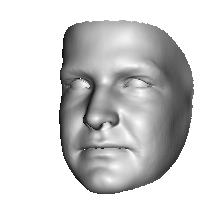} &
    \includegraphics[align=c,width=1.8cm]{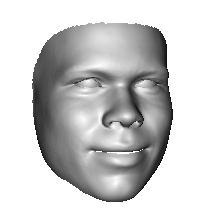} &
    \includegraphics[align=c,width=1.8cm]{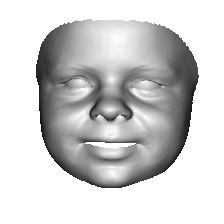} &
    \includegraphics[align=c,width=1.8cm]{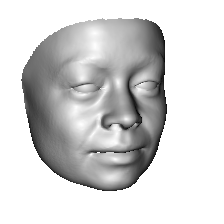} &
    \includegraphics[align=c,width=1.8cm]{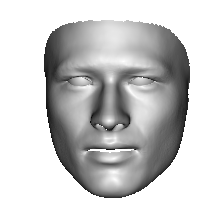} &
    \includegraphics[align=c,width=1.8cm]{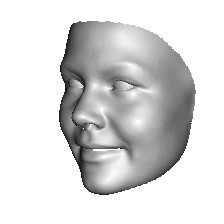}
    \\
    \small &$15.52\%$ & $12.93\%$ & $5.17\%$ & $3.45\%$ & $2.59\%$ & $2.59\%$ & $1.72\%$ & $0.86\%$ & $0.86\%$ \\
(e) &
    \includegraphics[align=c,width=1.8cm]{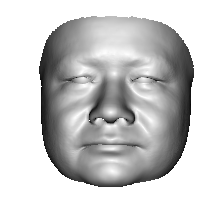} &     \includegraphics[align=c,width=1.8cm]{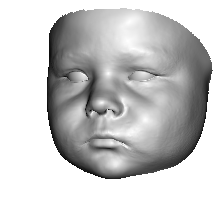} &
    \includegraphics[align=c,width=1.8cm]{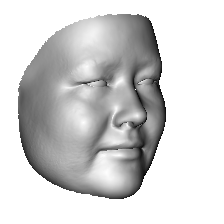} &
    \includegraphics[align=c,width=1.8cm]{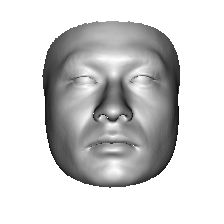} &
    \includegraphics[align=c,width=1.8cm]{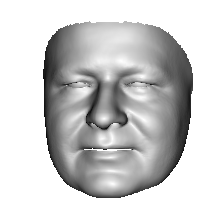} &
    \includegraphics[align=c,width=1.8cm]{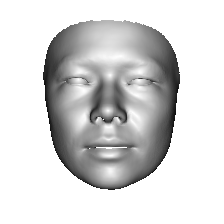} &
    &
   &
    \\
    \small &$32.76\%$ & $3.45\%$ & $3.45\%$ & $2.59\%$ & $2.59\%$ & $0.86\%$ &  &  &   \\
   (f) &
    \includegraphics[align=c,width=1.8cm]{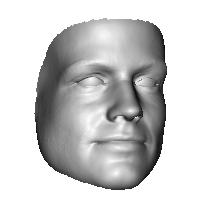} &     \includegraphics[align=c,width=1.8cm]{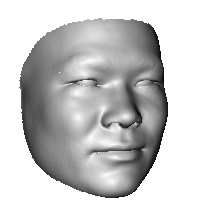} &
    \includegraphics[align=c,width=1.8cm]{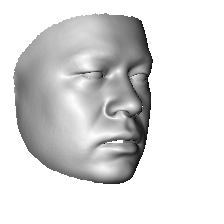} &
    \includegraphics[align=c,width=1.8cm]{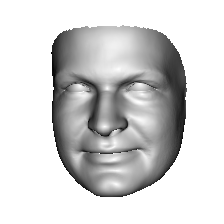} &
    \includegraphics[align=c,width=1.8cm]{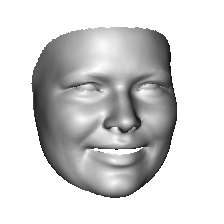} &
    \includegraphics[align=c,width=1.8cm]{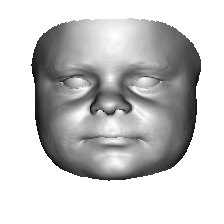} &
    \includegraphics[align=c,width=1.8cm]{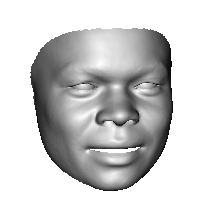} &
    \includegraphics[align=c,width=1.8cm]{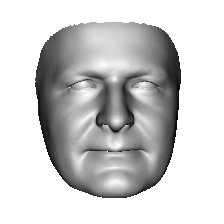} &
    \includegraphics[align=c,width=1.8cm]{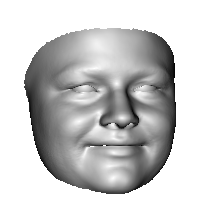}
    \\
    \small &$18.97\%$ & $11.21\%$ & $5.17\%$ & $5.17\%$ & $2.59\%$ & $2.59\%$ & $2.59\%$ & $1.72\%$ & $0.86\%$ \\
    \end{tabular}
\caption{Set of 3D master face images generated with each of the Coverage Search methods: LM-MA-ES Coverage Search on clustered data (a,b), greedy-Coverage Search methods (c,d) LM-MA-ES, (e,f) LM-MA-ES+Success Predictor. Embedded using either (a,c,e) Kim et al. or (b,d,f) FR3DNet face descriptors. Below each image, its MSC score is given.}
\label{fig:cover_3d_image}
\end{figure*}

{Fig.~\ref{fig:cover_3d_image} presents the generated 3D master face images for each of the experiments listed in Table {\ref{table:ds_3d_cover}}. The 3D master faces are arranged from left to right in descending order of their MSC values, where the MSC reported below each image excludes faces already covered by previous master faces. For some experiments, less than nine master faces are presented - these are the cases in which some iterations did not succeed in covering faces that remained uncovered after previous iterations.}.

\begin{table}
\caption{Percentage of 3D dataset covered by generated 3D master faces}
\label{table:ds_3d_cover}
\begin{center}
\begin{tabular}{lcc}
\toprule
 & Kim et al. & FR3DNet  \\
\midrule
\makecell[l]{Coverage Search LM-MA-ES\\on clustered data}& $37.93\%$ &  $36.21\%$ \\
\hline
\makecell[l]{$greedy$-Coverage Search\\ LM-MA-ES}& $41.38\%$ &  $45.69\%$ \\
\hline
\makecell[l]{$greedy$-Coverage Search\\ LM-MA-ES+Success Pred.}& $\mathbf{45.69}\%$&  $\mathbf{50.86}\%$  \\ 
\bottomrule
\end{tabular}
\end{center}
\vspace{-4mm}
\end{table}

\section{Experimnets - Combined 2D and 3D}
\label{subsec:2d3d_scenario}
We next evaluate our method in a combined 2D and 3D scenario, where the face recognition model receives as input both the 2D RGB face image and the corresponding 3D face structure. In particular, FaceNet \cite{fd:Facenet} (trained on CasiaWebface~\cite{ds:casia}) and FR3DNet \cite{FaceRec3D} are used as the 2D and 3D recognition models, respectively. A given pair of corresponding 2D and 3D faces must pass both the 2D and 3D recognition models in order to be authenticated. The Texas 3D dataset contains pairs of high-resolution and aligned color and range images, supporting this experiment.

Each model of the 2D and 3D recognition models requires its own recognition threshold, i.e. two thresholds are defined in total. We therefore perform a two-dimensional grid search in the square $[0,1]\times[0,1]$ to find a pair of thresholds for which the FAR of the combined model on the entire Texas 3D dataset is as close as possible to its FRR. In this case, a pair of thresholds was found that leads to an EER of 0.004

As the generator $\G$, we use StyleGAN2 \cite{Karras2019stylegan2} followed by Deng et al. \cite{deng2019accurate}, similarly to Sec. \ref{subsec:3d_scenario}, but in this scenario the generated 2D RGB output of StyleGAN2 is used for authentication in addition to the 3D Reconstruction. Moreover, the mask predicted by the 3D reconstruction model is used to extract the facial region of interest of the 2D RGB image precisely prior to inputting it to the 2D face recognition model.

\subsection{Dataset Coverage} In this section, we seek to generate at most nine paired 2D and 3D master faces, which cover the maximal of dataset $D$ (Sec. \ref{subsubec:3d_coverage}), where $|D|=116$.

Table \ref{table:ds_2d3d_cover} lists the MSC results obtained by applying Alg. \ref{alg:greedy} with the LM-MA-ES evolutionary algorithm and our proposed LM-MA-ES+Success predictor. Clustering was not applied, since there are two different metrics. 

\begin{figure*}
\centering
\begin{tabular}{@{~}l@{~}c@{~}c@{~}c@{~}c@{~}c@{~}c@{~}c@{~}c@{~}c@{~}}
(a)&
    \includegraphics[align=c,width=1.8cm]{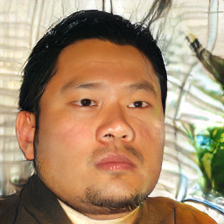} & 
    \includegraphics[align=c,width=1.8cm]{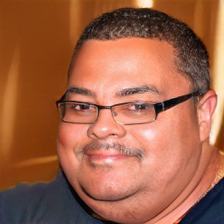} & 
    \includegraphics[align=c,width=1.8cm]{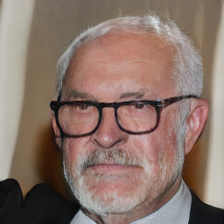} & 
    \includegraphics[align=c,width=1.8cm]{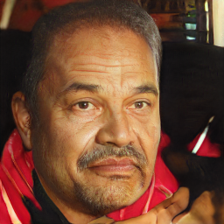} & 
    \includegraphics[align=c,width=1.8cm]{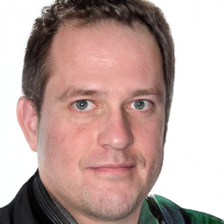} & 
    \includegraphics[align=c,width=1.8cm]{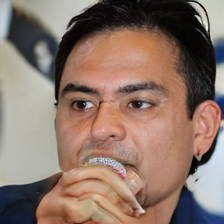} & 
    \includegraphics[align=c,width=1.8cm]{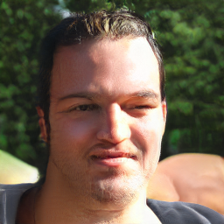} & 
    \includegraphics[align=c,width=1.8cm]{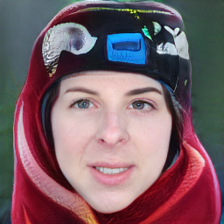} & 
    \includegraphics[align=c,width=1.8cm]{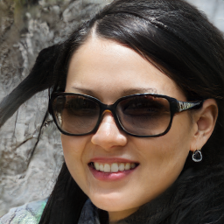} \\
    &
    \includegraphics[align=c,width=1.8cm]{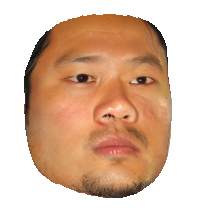} & 
    \includegraphics[align=c,width=1.8cm]{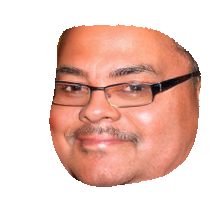} &     
    \includegraphics[align=c,width=1.8cm]{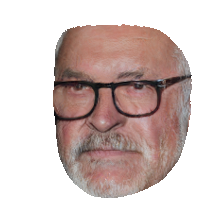} &     
    \includegraphics[align=c,width=1.8cm]{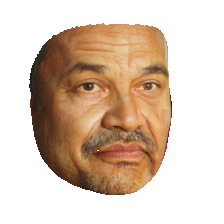} &     
    \includegraphics[align=c,width=1.8cm]{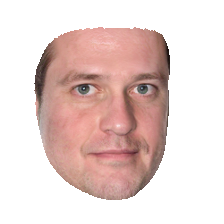} &     
    \includegraphics[align=c,width=1.8cm]{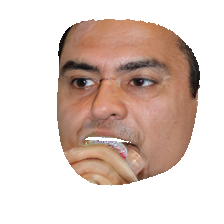} &     
    \includegraphics[align=c,width=1.8cm]{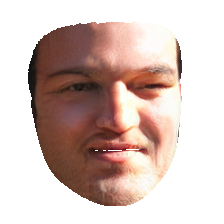} &     
    \includegraphics[align=c,width=1.8cm]{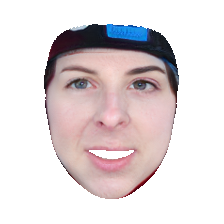} &     
    \includegraphics[align=c,width=1.8cm]{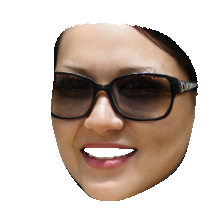} \\
    &
    \includegraphics[align=c,width=1.8cm]{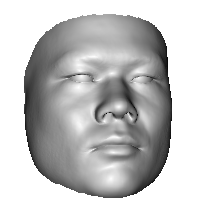} & 
    \includegraphics[align=c,width=1.8cm]{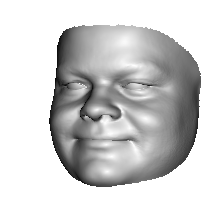} &     
    \includegraphics[align=c,width=1.8cm]{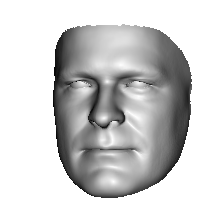} &     
    \includegraphics[align=c,width=1.8cm]{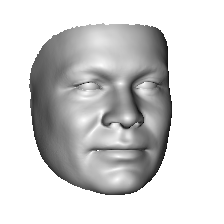} &     
    \includegraphics[align=c,width=1.8cm]{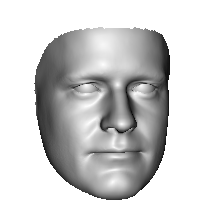} &     
    \includegraphics[align=c,width=1.8cm]{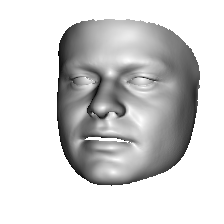} &     
    \includegraphics[align=c,width=1.8cm]{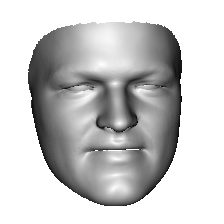} &     
    \includegraphics[align=c,width=1.8cm]{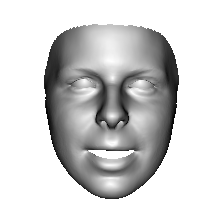} &     
    \includegraphics[align=c,width=1.8cm]{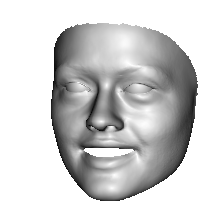} \\
    \small &$21.60\%$ & $10.30\%$ & $6.03\%$ & $4.31\%$ & $3.45\%$ & $3.45\%$ & $2.59\%$ & $2.59\%$ & $1.72\%$ \\
\midrule
(b)&
    \includegraphics[align=c,width=1.8cm]{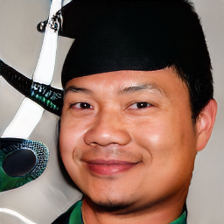} &   
    \includegraphics[align=c,width=1.8cm]{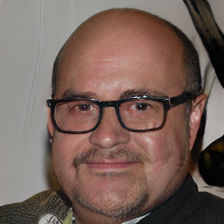} &   
    \includegraphics[align=c,width=1.8cm]{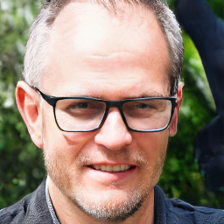} &   
    \includegraphics[align=c,width=1.8cm]{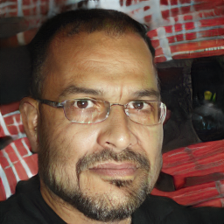} &   
    \includegraphics[align=c,width=1.8cm]{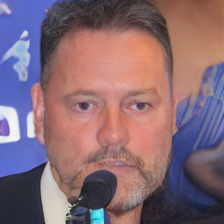} &   
    \includegraphics[align=c,width=1.8cm]{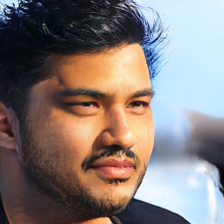} &   
    \includegraphics[align=c,width=1.8cm]{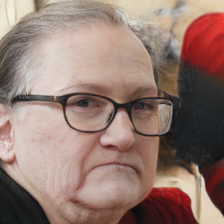} &   
    \includegraphics[align=c,width=1.8cm]{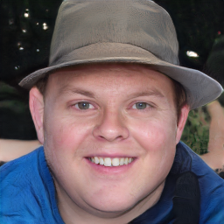} &   
    \includegraphics[align=c,width=1.8cm]{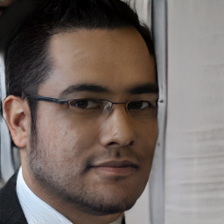} \\
    &
    \includegraphics[align=c,width=1.8cm]{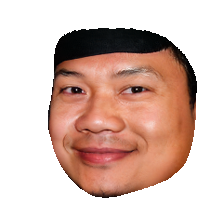} &     
    \includegraphics[align=c,width=1.8cm]{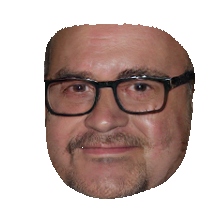} &     
    \includegraphics[align=c,width=1.8cm]{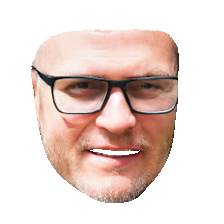} &     
    \includegraphics[align=c,width=1.8cm]{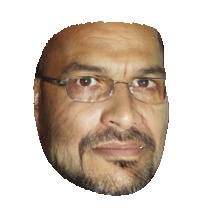} &     
    \includegraphics[align=c,width=1.8cm]{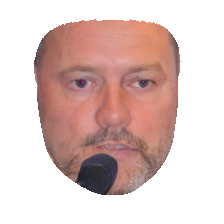} &     
    \includegraphics[align=c,width=1.8cm]{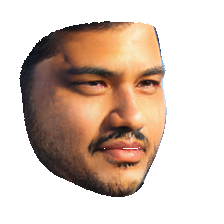} &     
    \includegraphics[align=c,width=1.8cm]{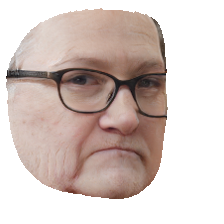} &     
    \includegraphics[align=c,width=1.8cm]{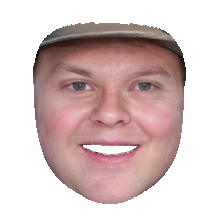} &     
    \includegraphics[align=c,width=1.8cm]{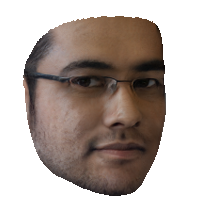} \\
    &
    \includegraphics[align=c,width=1.8cm]{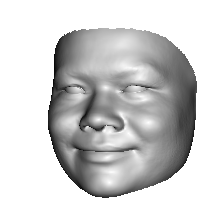} &     
    \includegraphics[align=c,width=1.8cm]{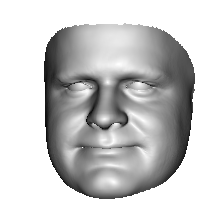} &     
    \includegraphics[align=c,width=1.8cm]{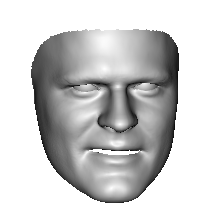} &     
    \includegraphics[align=c,width=1.8cm]{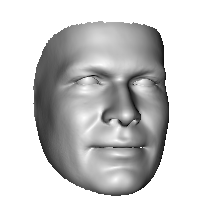} &     
    \includegraphics[align=c,width=1.8cm]{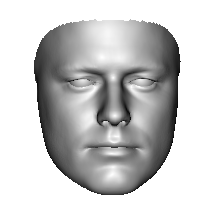} &     
    \includegraphics[align=c,width=1.8cm]{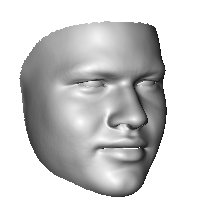} &     
    \includegraphics[align=c,width=1.8cm]{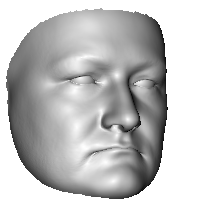} &     
    \includegraphics[align=c,width=1.8cm]{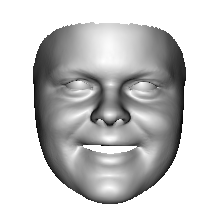} &     
    \includegraphics[align=c,width=1.8cm]{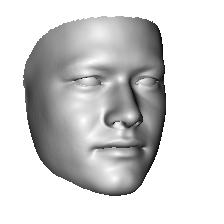} \\
    \small &$19.00\%$ & $10.30\%$ & $5.17\%$ & $5.17\%$ & $4.31\%$ & $4.31\%$ & $3.45\%$ & $3.45\%$ & $2.59\%$ \\

    \end{tabular}
\caption{Set of generated master faces in the combined 2D and 3D scenario, generated using the greedy-Coverage method by (a) LM-MA-ES or (b) LM-MA-ES+Success Predictor. For each, the 2D RGB master faces, the masked 2D face and the corresponding 3D master face are presented top to bottom. The MSC scores are listed below each master face. }\label{fig:cover_2d3d_image}
\end{figure*}

Fig. \ref{fig:cover_2d3d_image} presents the nine generated 2D master faces, their masked version and their corresponding 3D master faces. Evidently, both are able to cover a large portion of the dataset. While the single master face, which achieves the highest MSC, is obtained by the original variant of LM-MA-ES, our success predictor slightly improves the total coverage obtained by the set of nine generated master faces.

\begin{table}
\caption{Percentage of paired 2D and 3D dataset covered by generated master faces}
\label{table:ds_2d3d_cover}
\begin{center}
\begin{tabular}{lc}
\toprule
 & \makecell[c]{FaceNet (2D) \& \\ FR3DNet (3D)}  \\
\midrule
\makecell[l]{$greedy$-Coverage Search\\ LM-MA-ES}& $56.03\%$  \\
\hline
\makecell[l]{$greedy$-Coverage Search\\ LM-MA-ES+Success Pred.}& $\mathbf{57.76}\%$\\ 
\bottomrule
\end{tabular}
\end{center}
\vspace{-4mm}
\end{table}

\section{CONCLUSIONS AND FUTURE WORKS}

Our results imply that face-based authentication is vulnerable even when there is no information on the target identity. This is true for both 2D and 3D recognition methods.

Interestingly, the obtained 2D faces are not blurry and their pose is mostly frontal. The generated images on the LFW tend to be of older faces. Since face recognition methods are trained on faces representing different ages, the representation has some age invariance. It is possible that methods make use of this fact. However, we do not observe such a tendency toward facial hair or glasses. 

We further note that according to \cite{han2014age}, the group of 61+ years-old Caucasians is the third most common group in the LFW dataset, with groups of younger(21-60 years) Caucasians even more common. Nevertheless, the group of 61+ years-old Caucasian males is usually less varied, so a single older master face can cover a larger percentage of its group. With successive iterations of the coverage algorithm it covers less represented groups in the dataset, since each iteration is performed on the reduced dataset. Eventually master faces of several ethnicities and ages are generated. The lower number of female faces among the the nine master faces generated by our method on the LFW dataset corresponds to the much lower frequency of female faces (22\%) in the LFW dataset according to \cite{han2014age}. {The experiments on the RFW dataset, which is more racially-diverse than the LFW dataset, indeed demonstrate that our method is not limited in generating master faces of a specific ethnic group.}
In order to provide a more secure solution for face recognition systems, anti-spoofing~\cite{antispoofing_survey}  methods are usually applied. {For example, liveliness detection models~\cite{xu2021improving} detect if the presented face moves naturally, while image quality-based methods~\cite{wang2022patchnet,wang2022face} seek for artifacts, which are more common in artificial face images than in natural ones.} Our method could be combined with other existing methods to bypass such defenses. For example, DeepFake methods \cite{DeepFake_attack} can be used to animate the generated master faces and overcome liveliness detection methods.

{As examined by~\cite{nguyen2021master,terhorst2022limited}, the transferability of master faces is limited, especially when they are examined on a different face descriptor than the descriptor, which was used in the process of their generation. We leave the research of improving the transferability of master faces for future work.} 
Another interesting direction would be exploring the possibility of using master faces to help protect face recognition systems against dictionary attacks and reduce the overall false positive rate. Anecdotally - although, as far as we can ascertain, unreported in publications - the same set of faces appear multiple times among false matches.

\section{ACKNOWLEDGMENTS}
This project has received funding from the European Research Council (ERC) under the European Union's Horizon 2020 research and innovation programme (grant ERC CoG 725974). This research was partially
supported by The Yandex Initiative for Machine Learning. We thank Dr. Syed Zulqarnain Gilani and Dr. Donghyun Kim, for providing their 3D face recognition models.



%





\ifCLASSOPTIONcaptionsoff
  \newpage
\fi



\bibliographystyle{IEEEtran}
\bibliography{IEEEabrv, tbiom}
%




%



\end{document}